\documentclass[a4paper,12pt]{article}

\usepackage{amssymb}
\usepackage{amsfonts}
\usepackage{amsmath}
\usepackage{latexsym}
\usepackage{graphicx,color}
\usepackage{caption}
\usepackage{graphicx}
\graphicspath{ {images/} }
\usepackage{subcaption}
\usepackage{mathrsfs}
\usepackage[colorlinks=true, allcolors=blue]{hyperref}

\newskip\humongous \humongous=0pt plus 1000pt minus 1000pt

\newif\ifdtup

\allowdisplaybreaks[1]

\catcode`\@=11

\@addtoreset{equation}{section}

\def\@normalsize{\@setsize\normalsize{15pt}\xiipt\@xiipt
\abovedisplayskip 14pt plus3pt minus3pt%
\belowdisplayskip \abovedisplayskip
\abovedisplayshortskip \z@ plus3pt%
\belowdisplayshortskip 7pt plus3.5pt minus0pt}

\def\small{\@setsize\small{13.6pt}\xipt\@xipt
\abovedisplayskip 13pt plus3pt minus3pt%
\belowdisplayskip \abovedisplayskip
\abovedisplayshortskip \z@ plus3pt%
\belowdisplayshortskip 7pt plus3.5pt minus0pt
\def\@listi{\parsep 4.5pt plus 2pt minus 1pt
     \itemsep \parsep
     \topsep 9pt plus 3pt minus 3pt}}

\relax

\catcode`@=12

\catcode`\@=11

\def\section{\@startsection{section}{1}{\z@}{3.5ex plus 1ex minus
   .2ex}{2.3ex plus .2ex}{\large\bf}}

\def\SymBoxes#1#2#3#4{\newdimen\un@t \un@t#3%
\raisebox{#1}{\rule{#2\un@t}{#4}\hskip-#2\un@t
\@tempdimb\un@t \advance\@tempdimb by-#4\@tempcntb#2\relax%
\@whilenum{\@tempcntb>0}\do{
\rule{#4}{\un@t}\hskip\@tempdimb \advance\@tempcntb by\m@ne}%
\hskip-#2\un@t \rule[\un@t]{#2\un@t}{#4}%
\rule[\un@t]{#4}{#4}\hskip-#4
\rule{#4}{\un@t}}\hskip-#4}                

\DeclareMathAlphabet{\pazocal}{OMS}{zplm}{m}{n}

\begin{document}

\newcommand{\beq}{\begin{equation}}
\newcommand{\eeq}{\end{equation}}
\newcommand{\bea}{\begin{eqnarray}}
\newcommand{\eea}{\end{eqnarray}}
\newcommand{\beas}{\begin{eqnarray*}}
\newcommand{\eeas}{\end{eqnarray*}}
\newcommand{\defi}{\stackrel{\rm def}{=}}
\newcommand{\non}{\nonumber}
\newcommand{\bquo}{\begin{quote}}
\newcommand{\enqu}{\end{quote}}
\renewcommand{\(}{\begin{equation}}
\renewcommand{\)}{\end{equation}}
\def \eqn#1#2{\begin{equation}#2\label{#1}\end{equation}}
\def\IZ{{\mathbb Z}}
\def\IR{{\mathbb R}}
\def\IC{{\mathbb C}}
\def\IQ{{\mathbb Q}}
\def\de{\partial}
\def\Tr{ \hbox{\rm Tr}}
\def\H{ \hbox{\rm H}}
\def\HE{ \hbox{$\rm H^{even}$}}
\def\HO{ \hbox{$\rm H^{odd}$}}
\def\K{ \hbox{\rm K}}
\def\Im{ \hbox{\rm Im}}
\def\Ker{ \hbox{\rm Ker}}
\def\const{\hbox {\rm const.}}
\def\o{\over}
\def\im{\hbox{\rm Im}}
\def\re{\hbox{\rm Re}}
\def\bra{\langle}\def\ket{\rangle}
\def\Arg{\hbox {\rm Arg}}
\def\Re{\hbox {\rm Re}}
\def\Im{\hbox {\rm Im}}
\def\exo{\hbox {\rm exp}}
\def\diag{\hbox{\rm diag}}
\def\longvert{{\rule[-2mm]{0.1mm}{7mm}}\,}
\def\a{\alpha}
\def\dag{{}^{\dagger}}
\def\tq{{\widetilde q}}
\def\p{{}^{\prime}}
\def\W{W}
\def\N{{\cal N}}
\def\hsp{,\hspace{.7cm}}

\def\br{\nonumber\\}
\def\IZ{{\mathbb Z}}
\def\IR{{\mathbb R}}
\def\IC{{\mathbb C}}
\def\IQ{{\mathbb Q}}
\def\IP{{\mathbb P}}
\def \eqn#1#2{\begin{equation}#2\label{#1}\end{equation}}

\newcommand{\sgm}[1]{\sigma_{#1}}
\newcommand{\idd}{\mathbf{1}}

\newcommand{\C}{\ensuremath{\mathbb C}}
\newcommand{\Z}{\ensuremath{\mathbb Z}}
\newcommand{\R}{\ensuremath{\mathbb R}}
\newcommand{\rp}{\ensuremath{\mathbb {RP}}}
\newcommand{\cp}{\ensuremath{\mathbb {CP}}}
\newcommand{\vac}{\ensuremath{|0\rangle}}
\newcommand{\vact}{\ensuremath{|00\rangle}                    }
\newcommand{\oc}{\ensuremath{\overline{c}}}

\newcommand{\scrip}{\mathscr{I}^{+}}
\newcommand{\scrim}{\mathscr{I}^{-}}
\newcommand{\scri}{\mathscr{I}}

\begin{titlepage}
\def\thefootnote{\fnsymbol{footnote}}


\begin{center}
{\large
{\bf A Holographic Entanglement Entropy at Spi 
}
}
\end{center}

\begin{center}
{Abir Ghosh$^a$\footnote{\texttt{abirghosh.physics@gmail.com}}, \ \ \ Chethan Krishnan$^a$\footnote{\texttt{chethan.krishnan@gmail.com} }}


\end{center}

\renewcommand{\thefootnote}{\arabic{footnote}}

\begin{center}
\vspace{-0.2cm}
$^a$ {Center for High Energy Physics,\\
Indian Institute of Science, Bangalore 560012, India}

\end{center}

\noindent
\begin{center} {\bf Abstract} \end{center}
Defining finite entanglement entropy for a subregion in quantum field theory requires the introduction of two logically independent scales: an IR scale that controls the size of the subregion, and a UV cut-off. In AdS/CFT, the IR scale is the AdS lengthscale, the UV cut-off is the bulk radial cut-off, and the subregion is specified by dimensionless angles. This is the data that determines Ryu-Takayanagi surfaces and their areas in AdS/CFT. We argue that in asymptotically flat space there exists the notion of a ``spi-subregion" that one can associate to  spatial infinity (spi). Even though geometrically quite different from an AdS subregion, this angle data has the crucial feature that it allows an interpretation as a bi-partitioning of spi. Therefore, the area of the RT surface associated to the spi-subregion can be interpreted as the entanglement entropy of the reduced density matrix of the bulk state under this bi-partition, as in AdS/CFT. For symmetric spi-subregions, these RT surfaces are the waists of Asymptotic Causal Diamonds. In empty flat space they reduce to Rindler horizons, and are analogues of the AdS-Rindler horizons of Casini, Huerta \& Myers. We connect these results to previous work on minimal surfaces anchored to screens in empty space, but also generalize the discussion to the case where there are black holes in the bulk. The phases of black hole RT surfaces as the spi-subregion is varied, naturally connect with those of black holes (small and large) in AdS. A key observation is that the radial cut-off is associated to an IR scale in flat space -- and in fact there are no UV divergences. We argue that this is consistent with previous suggestions that in sub-AdS scales the holographic duality is an IR/IR correspondence and that the degrees of freedom are {\em not} those of a local QFT, but those of long strings. Strings are of course, famously UV finite.

\vspace{1.6 cm}
\vfill

\end{titlepage}

\setcounter{page}{2}
\tableofcontents

\setcounter{footnote}{0}

\section{Introduction}

A remarkable fact about the AdS/CFT correspondence is that the holographer has a ``place to stand'' \cite{Witten, Polchinski}. The conformal boundary of (Poincare) AdS is Minkowski space, and therefore in many ways it is the best setting one could hope for, to place a dual theory: the boundary almost begs for the theory to be a local quantum field theory. In flat space on the other hand, even though the conformal boundary is certainly well-defined, it has no causal structure. Indeed the metric seems to lose {\em two} dimensions at the null boundary. Since most theories we are familiar with are local (and often Lorentzian), coming up with a theory that could live on the boundary of an asymptotically flat spacetime is a challenge. 

Even though understanding dynamics fully is likely to take more ideas, understanding the entanglement structure of the flat space hologram may be more tractable -- it is natural to suspect that entanglement entropy (EE) has more to do with spatial infinity $spi$, rather than null or timelike infinity. Indeed spi is where bulk spatial slices end, and therefore one expects it to carry a copy of the holographic Hilbert space at each instant of time\footnote{In the bulk of both flat space and AdS, evolution happens via spatial slices that evolve in time. Since EE is a quantity that can be defined at any instant of time in quantum mechanical theories, one would like to associate it to suitably defined asymptotic data on a Cauchy slice. Indeed, the Ryu-Takayanagi formula \cite{RT1, RT2} is a realization of that expectation.}. This ties in with the observation that in flat space general relativity, charges are  naturally defined either directly at spi in the Ashtekar-Hansen gauge \cite{AshtekarHansen, AshtekarRomano} or at the past boundary of future null infinity $\scrip_-$ (which is adjacent to spi) in the Bondi \cite{StromingerReview} and Special Double Null \cite{SDN1,SDN2,SDN3} gauges. Motivated by these observations, in this paper we will define and explore a holographic entanglement entropy that is naturally tied to spi, by introducing the notion of a spi-subregion. We will see that previous work on holographic entanglement entropy in empty Minkowski space defined on screens \cite{TakayanagiFlat, Qi, Chethan_1}, has a natural understanding in terms of our construction, via an IR scale. Our slightly more abstract definition will allow us to generalize these calculations to asymptotically flat spacetimes containing black holes. We find RT phase transitions in these spacetimes that naturally connect with those of small and large AdS black holes (see eg., \cite{Freivogel}) as spi-subregion ``sizes'' are varied. 

The biggest difference\footnote{We will be more precise later.} between a spi-subregion and an ordinary subregion is that one needs to specify a certain asymptotic ``approach'' angle, $\theta_{asymp}$, to define a spi-subregion\footnote{The approach angle has a natural ordering, and allows a (bulk-motivated) notion of inclusion of symmetric spi-subregions on the boundary. This allows us to meaningfully retain a notion of ``size" of the symmetric spi-subregion. But we emphasize that this should not be conflated mindlessly with the size of an ordinary subregion.}.  Spi is a description of spatial infinity that involves ``ultimate'' \cite{Gibbons} directions and accelerations, and that explains our choice of neologism.  But it should be emphasized that there are a couple of different descriptions of spatial infinity in the market \cite{AshtekarHansen, AshtekarRomano, Friedrich1, Friedrich2, SpiContrast}, and that the definition of spi often includes structures to clarify the asymptotic solution of Einstein's equations. We will not need these details because our interest is in the {\em kinematical} structure of the conformal boundary on a {\em spatial} slice. A similar statement applies in AdS as well -- in order to define an ordinary subregion on the conformal boundary of AdS, we do not need to specify fall-off conditions\footnote{Other than the general understanding that there is {\em some} notion of an asymptotically AdS spacetime that breathes life into quantum gravity in AdS. Similar statements hold in flat space as well.} or solve Einstein's equations asymptotically. Any reasonable definition of spatial infinity that incorporates the conformal boundary structure of Minkowski space will be sufficient for our purposes. So in this paper, we will use the phrase ``spi'' to refer to this minimal structure.

We believe that the structural differences we see in spi-subregions, are closely related to the lack of locality in the hologram of flat space. To explain this, let us first review some facts. In order to define a finite entanglement entropy for subregions in quantum field theory \cite{CasiniQFT}, we need to introduce two length scales. The first is an IR scale, which captures the subregion size. The second is a UV cut-off, which is necessary to regulate the short-distance divergences of local QFT. This UV cut-off is not merely a technicality, but related ultimately to the fact that entanglement is naturally associated to algebras of observables and not to states in QFT \cite{WittenEntanglement, WittenAlgebras}.

A basic fact about holographic entanglement entropy in AdS/CFT \cite{RT1,RT2} is that it requires the introduction of only one of these scales. This is related to the fact that CFT does not have an intrinsic scale, and the IR scale associated to the subregion size is in effect taken to be the AdS length scale $L$. In global AdS$_{d+1}$, a spherical subregion on the boundary $S^{d-1}$ is defined by a dimensionless solid $S^{d-2}$-ball determined by a point on $S^{d-1}$ (the center of the ball) and the polar angle that captures the (angular) radius of the ball. Similarly in Poincare AdS, the boundary Minkowski spacetime metric is built of dimensionless coordinates except for the overall dimensions provided by the AdS length scale. More general subregions require more complicated data, but this data is again dimensionless.

We will see that a loosely similar situation arises in asymptotically flat spaces as well, but with some conceptually crucial differences. Just as in AdS, in order to define a finite (holographic) entanglement entropy, we need a bulk IR cutoff in flat space. But the (bulk) areas of the associated RT surfaces scale as volumes on the screen, and not as areas \cite{TakayanagiFlat, Qi,  Chethan_1}. Since they are extensive on the screen and diverge as the screen radius goes to infinity, they are naturally viewed as IR divergences. A second feature is that these IR divergences are the only divergences in the entanglement entropy, there are no other (in particular no UV) divergences. It has been suggested that the holographic duality is an IR/IR duality in flat space, and that the degrees of freedom are not those of a local QFT, but those of long strings \cite{Verlinde}. The non-local interactions associated to the long strings are consistent both with the volume scaling as well as the absence of UV divergences that we find. 

We will be dealing with classical bulk EFT in this paper, as in the case of the EE computations in the original Ryu-Takayanagi papers \cite{RT1, RT2, HRT}. Of course, one could consider quantum corrections in the bulk. It would then become natural to have divergences in the bulk EE which can be interpreted as corrections to the bulk Newton's constant \cite{SusskindUglum}. These divergences show up via EE of {\em bulk} fields in the EFT across the RT surface. Computing this in AdS can be done using the quantum extremal surface (QES) prescription \cite{EngWall}. Since it is a bulk prescription, it has natural adaptations to flat space as well\footnote{See \cite{Chethan_2} for some discussions on this.}. When we talk about UV/IR  divergences in this paper we mean divergences in the holographic dual theory -- in the bulk they translate to divergences due to the unbounded RT surface and its area. This should not be confused with the divergences in Newton's constant which are cured by the QES prescription due to entanglement of bulk fields across the RT surface. This latter aspect should work largely analogously in both AdS and flat spaces, so our focus in this paper will be on the features that are special to flat space.

The screens in flat space on which previous definitions of holographic EE were made \cite{Qi, TakayanagiFlat, Chethan_1} become more comprehensible when viewed as regulators of EE for spi-subregions, by making the parallels and distinctions with AdS clearer. In AdS, since the dual description is a CFT, there is no scale associated to the ``size" of a subregion. Instead one has a cut-off that regulates the UV divergences of the CFT. The opposite is true in flat space, where the required cut-off is (presumably) in the length of the long string, and there is no regulator needed for the UV because of the UV-finiteness of strings. This perspective we feel, underscores the relationship between the screen in flat space vs. the screen in AdS, and the fact that we are dealing with a sort of ``symmetry'' between the two holographic theories. Holographic correspondence \cite{BudhadityaGeneral, VyshnavVestige} and black holes \cite{BalaHairyBH} have been studied previously in flat space with screens, and striking similarities have been noted with AdS/CFT.

The main goal of the first part of this paper is to build enough background in both AdS and flat space to clarify the notion of a spi-subregion. The AdS discussion is to be viewed as a biased review that emphasizes the elements that will be useful in our  comparisons with flat space -- even a reader familiar with AdS RT surfaces may find it useful to skim through this discussion because some of the details we need maybe unfamiliar. Once the AdS discussion is in place as a relatively familiar foil, the flat space discussion can proceed (hopefully) painlessly. One of our key observations is that there is enough structure on spi to define a $d$-parameter family of bi-partitions of spi in a bulk spacetime that is $d+1$ dimensional. This data should be compared to the data required to specify a spherical subregion on the boundary of global AdS -- the ``center" of the spherical subregion and the ``size" of the subregion. In asymptotically flat space, the situation is quite different structurally, but there still exists the notion of a symmetric spi-subregion that is the analogue of a spherical subregion. We describe this later, but the impatient reader can glance at Section \ref{spisub} and figures \ref{fig:acd}, \ref{fig:acd_waist} and \ref{fig:re_cast}. 
Remarkably, it turns out that we can generalize this further and in fact come up with the notion of a more general spi-subregion when $d > 2$.  The fact that they exist only in $d > 2$ should be compared to the fact that in AdS$_3$ all boundary subregions are ``spherical" subregions while in higher dimensional AdS$_{d+1}$ there are more complicated possibilities. The distinction between a spi-subregion and an ordinary subregion also results in some crucial differences in their unions and intersections.

Once the notion of a spi-subregion is well-defined, a pragmatic question we will address is that of RT surfaces in asymptotically flat spaces that contain black holes. Here again, viewing the screen as a regulator (and only indirectly as a box for the black hole) is cleaner, because the RT surfaces in the geometry are again determined by asymptotic data (ie., the spi-subregion) and the black hole. When one is fixing boundary conditions for black hole RT surfaces on the screen on the other hand, it is less clear whether they have invariant meaning as holographic data. Here instead we get to view the screen radius simply as a regulator for the area functional. Perhaps for these reasons, while RT surfaces for empty flat space with screens have been studied in the literature, they have not been studied in the setting of flat space black holes. We do this in some detail here. The problem involves two pieces of dimensionless data - the ratio of the horizon radius to the screen radius, and the spi-subregion. We will identify the phase transitions in RT surfaces as we vary this data and will see a close parallel with RT surface phase transitions in small and large AdS black holes. 

This paper can be viewed as another step in the direction initiated in \cite{Chethan_1} and developed further in \cite{Chethan_2}. The discussion here is self-contained.  A related comment is that even though background subtraction for entropies in flat space has been around since the famous paper of \cite{GibbonsHawking}, there are multiple hints \cite{Chethan_2} that this may require a better understanding. See eg. footnote on p. 71 of \cite{Hartman}. In this paper, we find it most natural to interpret the screen as an IR scale, which is natural from the perspective advocated in \cite{Verlinde}. A v2 of \cite{Chethan_2} that incorporates some of the refinements we have learnt here and strengthens the manifesto presented there, is in the pipeline\footnote{Let us note that the Page curve discussion of \cite{Chethan_2} is insensitive to these nuances, because islands are not anchored to the screen.}. 




\section{Ryu-Takayanagi Data in AdS}\label{sec:bound_sub}

We start by defining a spherical subregion on the boundary of AdS. In global AdS we can place it around the North pole without loss of generality. With this choice, the subregion is characterized by the polar angle $\theta= \theta_\infty= {\rm const.}$ The boundary subregion is a codimension-2 surface in AdS$_{d+1}$.

In suitably defined Poincare coordinates (see Appendix, and in particular \eqref{eq:glob_co} and \eqref{eq:poin_co}) we can write
\begin{equation}
    \frac{\Sigma_{i=2}^{d}X_i^2}{X_1^2}=\tan^2{(\theta)}=\left[\frac{R^2-x_1^2}{x_1^2}+\frac{1}{4}\left(\frac{L}{x_1}-\frac{z^2}{Lx}-\frac{(R^2-t^2)}{Lx}\right)^2\right]
\end{equation}
where the first expression defines the angle in terms of AdS embedding coordinates. Here $R^2 \equiv \Sigma_{i=1}^{d-1}x_i^2$. In the limit $z\rightarrow0$, setting $t=0$ and $\theta \equiv \theta_\infty $ we get,
\begin{align}
    \tan^2{(\theta_{\infty})}&=\left[\frac{R^2-x_1^2}{x_1^2}+\frac{1}{4}\left(\frac{L}{x_1}-\frac{R^2}{L x_1}\right)^2\right]\nonumber\\
    \Rightarrow \quad\quad \sec{(\theta_{\infty})}&=\frac{1}{2}\left(\frac{L}{x_1}+\frac{R^2}{L x_1}\right)\label{eq:boundary_sphere_1}
\end{align}
We can re-write the above equation as 
\begin{equation}
    \left(x_1-L\sec(\theta_\infty)\right)^2+\Sigma_{i=2}^{d-1}x_i^2=L^2\tan^2{(\theta_\infty)}\label{eq:boundary_sphere_2}
\end{equation}
which is a $S^{d-2}$ sphere on the boundary with radius $R=L\tan{(\theta_\infty)}$ with its centre at $x_1=L\sec(\theta_\infty)$. This is the spherical subregion on Poincare boundary that maps to the $\theta= \theta_\infty$ spherical subregion on the global AdS boundary. Note that this sphere is centred off the origin because of the relative orientation of the Poincare and global coordinates. With the conventional definition of coordinates as in \eqref{eq:glob_co} and \eqref{eq:poin_co} the origin of the boundary Minkowski geometry lies along the $X_d$ axis in global coordinates. We can center the sphere at origin by exchanging $X_1$ and $X_d$ in \eqref{eq:poin_co}. This aligns the origin of the boundary Minkowski geometry with the north pole in global AdS. Then the analogue of \eqref{eq:boundary_sphere_1} is 
\begin{equation}
    2\cot{(\theta)}=\frac{L}{R}-\frac{R}{L}
\end{equation}
On solving for $R$ we get $R = L (\csc{(\theta_\infty)}-\cot{(\theta_\infty)})$.  This spherical subregion can be compared directly with the subregion discussed in \cite{Casini}. A quantity $\beta$ was defined in \cite{Casini} which is related to our quantities via
\begin{equation}
    R/L\equiv e^{-\beta} = \csc{(\theta_\infty)}-\cot{(\theta_\infty)}
\end{equation}

As pointed out in the introduction, we see that the dimensionless angle $\theta_\infty$ uniquely fixes the boundary spherical subregion upto the choice of the centre of the sphere which can be decided by setting the orientation of the north-pole in global coordinates. The IR scale associated to the size of the subregion is given by the AdS length scale $L$. This is true both in global AdS as well as Poincare. Of course these observations are natural, because the conformal field theory does not have an intrinsic scale -- $L$ is essentially a book-keeping length scale at this semi-classical level of discussion. A more complete bulk understanding of holographic entanglement entropy in AdS/CFT will perhaps involve understanding it better at finite $N$ and 't Hooft coupling $\lambda$, but our discussions here are limited to the semi-classical limit. 

In AdS/CFT one can also consider non-spherical boundary subregions which can be viewed as (if necessary, infinite) unions of spherical subregions. This is a consequence of the fact that the dual theory is a QFT. While the union structure is substantially different, we will see that a notion of general spi-subregions can be defined in asymptotically flat space as well. In any case, these can all be characterized by dimensionless data, which is our key point here.


\section{Ryu-Takayanagi Surfaces in AdS}

Once we have a subregion on the boundary, we can calculate the entanglement entropy of the subregion using the Ryu-Takayanagi prescription. This states (roughly) that the entanglement entropy of the subregion can be calculated as the area of a co-dimension 2 minimal bulk surface anchored on the boundary of the subregion. Explicit calculation of RT surfaces and their associated areas is often complicated, and can be done only numerically. Often in calculations in the literature, some form of background subtraction is implemented so that the result is finite and cut-off independent. But it should be emphasized that both the CFT quantity we are after (namely, entanglement entropy) and the bulk Ryu-Takayanagi prescription for computing it, are in fact cut-off dependent. So we will present some details which are often not emphasized in the AdS calculations, in a way that can be adapted reasonably straightforwardly to asymptotically flat space, while clarifying the similarities and distinctions.

\subsection{Empty AdS}

In empty AdS, the RT surfaces for spherical boundary subregions are simply horizons of topological black holes anchored to the subregion \cite{Casini}. Note that horizons are minimal surfaces. Topological black holes are nothing but AdS-Rindler wedges and are discussed in a language that connects with the boundary flat space, in Appendix \ref{AppA}. More generally, in {\em asymptotically} AdS spacetimes these can be viewed as waists of asymptotic causal diamonds (ACDs) in the language of \cite{Chethan_1}. In empty AdS they reduce to AdS-Rindler wedges. The advantage of the ACD language is that it is amenable to generalization, and applies even when the bulk is not empty, has black holes, is flat instead of AdS, etc.  We will see that it also connects naturally with the spi-subregion description of RT data in asymptotically flat spaces.
\begin{figure}[h]
\centering
     \begin{subfigure}[t]{0.45\textwidth}
            \centering
            \includegraphics[width=7.5cm]{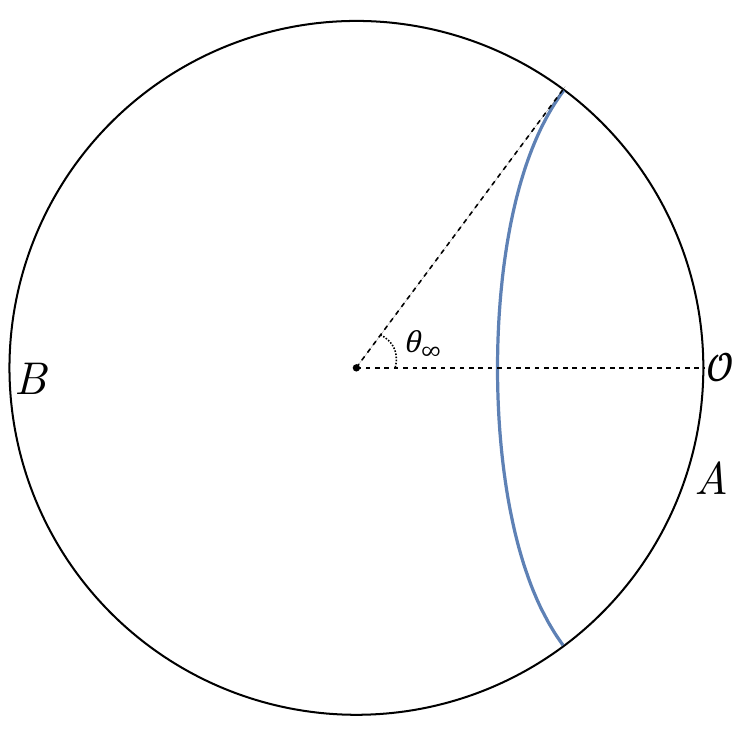}
            \caption{}
            \label{fig:ads_waist_1}
     \end{subfigure}
     \hfill
      \begin{subfigure}[t]{0.5\textwidth}
            \centering
            \includegraphics[width=7.5cm]{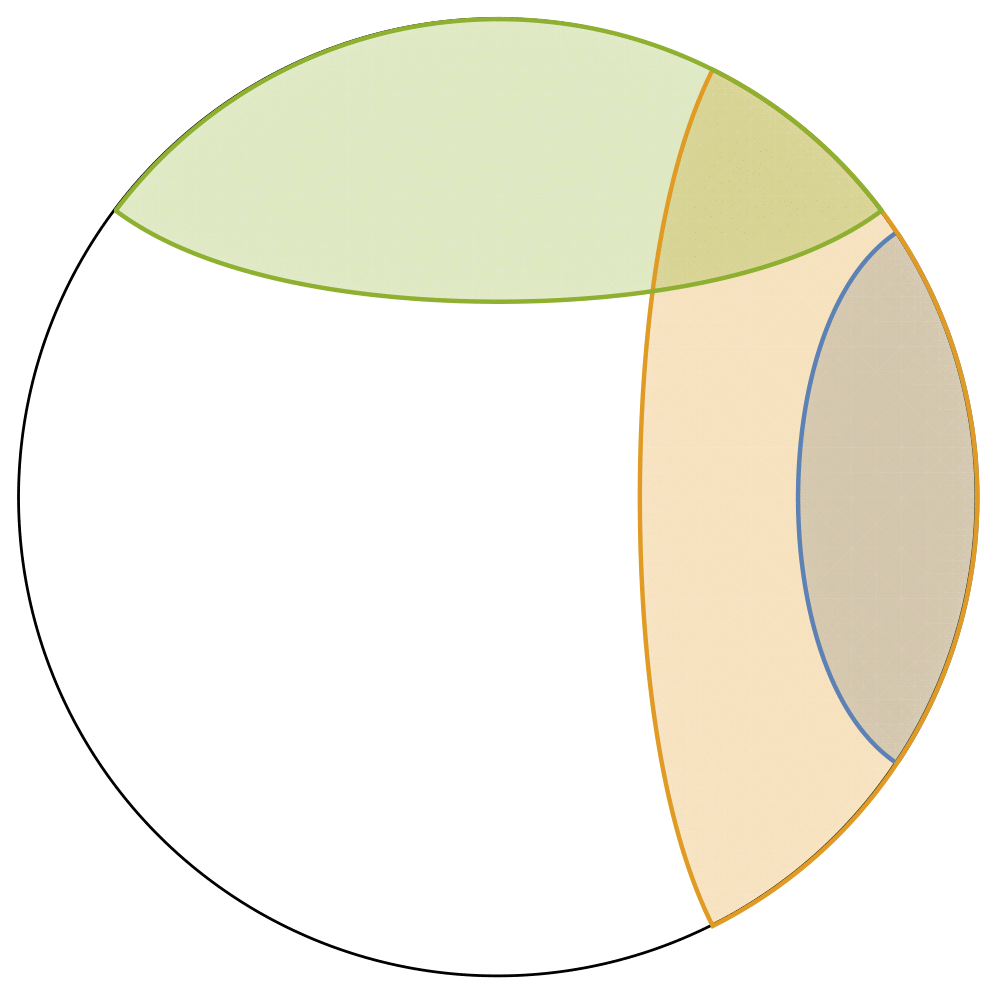}
            \caption{}
            \label{fig:ads_waist_2}
     \end{subfigure}  
     \caption{(a) The blue line is the RT surface associated with the subregion $A$ (or equivalently $B$). As pointed out in section \ref{sec:bound_sub}, after the alignment of axes the North pole lines up with the origin $\mathcal{O}$ of the boundary geometry. (b) The plot shows different RT surface for different subregions. The shaded regions are the regions inside the causal wedges associated with the subregions.}
     \label{fig:ads_waist}
\end{figure}

For empty AdS, the boundary theory is in the vacuum, a pure state. Therefore, the entanglement entropy of the subregion and its complement is the same. This feature of entanglement entropy is captured by the RT surfaces -- the subregion $A$ and it complement $\bar{A}$ have the same RT surface. Also, the entanglement entropy associated with the full spatial slice of the CFT is zero, which is again given by a trivial RT-surface in the limit $A \rightarrow0$. The above features are a consequence of the fact that an RT surface implements a bi-partition of the bulk, as well as the boundary. In flat space, when our goal is to construct an RT prescription at spi, we will see that these bi-partition properties are true there as well.

\subsubsection{AdS$_3$}\label{AdS_3}

We start with global AdS$_3$ where we can easily find analytic solutions. The AdS$_3$ metric is, 
\begin{equation}    
    ds^2=-\left(1+\frac{r^2}{L^2}\right)dt^2+\left(1+\frac{r^2}{L^2}\right)^{-1}dr^2+r^2d\theta^2
\end{equation}
Note that the angle $\theta$ here denotes a periodic coordinate with period $2\pi$, in higher dimensions we will use $\theta$ to denote the polar angle which has range from $0$ to $\pi$. To calculate the RT-surface we need to extremize the area functional of a curve defined by $r(\theta)$ or equivalently $\theta(r)$:
\begin{equation}\label{eq:area_func}
    \mathcal{A}=\int\sqrt{\left(1+\frac{r^2}{L^2}\right)^{-1}r'^2+r^2}\;d\theta=\int\sqrt{\left(1+\frac{r^2}{L^2}\right)^{-1}+r^2\theta'^2}\;dr
\end{equation}
Even though the functionals are equivalent in both $r(\theta)$ and $\theta(r)$ languages, the natural data for solving the resultant differential equations will change depending on the choice. Also, in higher dimensions since we do not have exact analytical solutions, the two languages give us equations that are more naturally solved in different regions of the spacetime. In the $r(\theta)$ language it is more natural to put integration data at the deepest point in the bulk, while the $\theta(r)$ language is more naturally adapted to a holographic perspective in that it is natural to think of the data at $r \rightarrow \infty$. But it should be noted that there is a correspondence between these two kinds of data -- this will be important sometimes when we work with the $r(\theta)$ language for convenience.

Since $\theta$ is a cyclic coordinate, the Euler-Lagrange equation becomes ($c$ is a constant of integration),
\begin{align}
    \quad\quad \frac{d\theta}{dr}&=\frac{cL}{r\sqrt{(r^2-c^2)(r^2+L^2)}} \label{eq:EL_eq}
\end{align}
Integrating \eqref{eq:EL_eq} and demanding $r=r_*$ for $\theta=0$ sets $c=r_*$. Then the final equation of the curve is,
\begin{equation}\label{eq:ads_3_sol}
    \theta =\tan^{-1}{\left(\frac{L}{r_*}\sqrt{\frac{r^2-r_*^2}{r^2+L^2}}\right)}
\end{equation}
where $r_*$ is the closest distance from from the origin and is related to $\theta_\infty\;$ as $\;\theta_\infty=\tan^{-1}(L/r_*)$ \cite{Freivogel}. This is an explicit realization of our previous statement that given a background, the data at the deepest point in the bulk can be used as a proxy for holographic RT data.

It turns out that (even though not emphasized in the literature) the above equation is in fact the universal solution associated to spherical subregions on the boundary in any number of dimensions for empty AdS (and not just AdS$_3$). A similar statement will be true in Minkowski space as well when dealing with symmetric spi-subregions -- the same equation will describe lines and (hyper-)planes in any dimension.

Asymptotically, $\theta_\infty$ captures the same (holographic) data as in section \ref{sec:bound_sub}. The integration constant we get on integrating \eqref{eq:EL_eq} is set to zero as it just rotates the curve about the origin. Substituting this solution into the area functional \eqref{eq:area_func} we get the entropy 
\begin{align}
    \mathcal{S}=\frac{\mathcal{A}}{4G}&=\frac{1}{4G}\;2L\ln{\left(\sqrt{r^2-r_*^2}+\sqrt{r^2+L^2}\right)}\Bigg|_{r_*} ^{1/\epsilon}\nonumber\\
    &=\frac{L}{2G} \ln{\left(\frac{2}{\epsilon\sqrt{r_*^2+L^2}}\right)} = \frac{c}{3}\ln{\left(\frac{2}{\epsilon L}\sin{\theta_\infty}\right)}
\end{align}
where $1/\epsilon$ is a large distance radial cut-off. With the coordinate redefinition $r=L \sinh{\rho}$ and using $c=3L/2G$, the final result is in exact agreement with the formulas in \cite{RT1, RT2} (see also \cite{Calabrese}) after a suitable renaming of the UV cut-off.

If we had treated $\theta$ as a polar angle (so that the $S^1$ is to be viewed as two points fibered on the $\theta$ segment) instead of as a cyclic coordinate, then the integration constant after \eqref{eq:EL_eq} integration would have to be treated more carefully. We cannot simply explain it away as the choice of origin of the cyclic coordinate. When $\theta$ is a polar angle, the integration constant is fixed by demanding regularity in the {\em bulk}.  This is useful to remember, because this is what happens in higher dimensions (as well as in asymptotically flat space) where we will work with a polar angle $\theta$.

\subsubsection{AdS$_4$}\label{sec:ads_4}

For AdS in higher dimensions an analytic solution when working in the $\theta(r)$ set up is not known to us, but we can still calculate the solution as a power series expansion. The AdS$_{d+1}$ metric is,
\begin{equation}
    ds^2=-\left(1+\frac{r^2}{L^2}\right)dt^2+\left(1+\frac{r^2}{L^2}\right)^{-1}dr^2+r^2d\Omega_{d-1}^2
\end{equation}
and for $d=3$ the area functional is\footnote{we have set $L=1$ for the purpose of calculation. Factors of $L$ can be restored dimensionally.},
\begin{equation} \label{eq:area_func_2}
    \mathcal{A}=2\pi \int \underbrace{r \sin(\theta (r)) \sqrt{\frac{1}{(r^2+1)}+r^2 \theta '(r)^2}}_{\equiv\ \mathcal{L}} \;dr
\end{equation}
Since the spherical subregion is symmetrical about north pole, we an take the RT surface to be also be symmetrical. From \eqref{eq:area_func_2} the minimal surface equation is
\begin{equation}\label{eq:EL_eq_2}
    \frac{\partial }{\partial \theta}\left(r \sin(\theta (r)) \sqrt{\frac{1}{(r^2+1)}+r^2 \theta '(r)^2}\right)=\frac{d}{d r}\frac{\partial }{\partial \theta'}\left(r \sin(\theta (r)) \sqrt{\frac{1}{(r^2+1)}+r^2 \theta '(r)^2}\right)
\end{equation}
Consider a power series expansion\footnote{Note that to match the dimensions here, we need to view the expansion as an expansion in the dimensionless radial coordinate $\frac{r}{L}$. We will emphasize an analogous feature when we work with spi-subregion data on the conformal boundary of flat space.} for large values of $r$,
\begin{equation}\label{eq:exp_1}
        \theta(r)=\sum_{i=0}\frac{\theta_i}{r^i}
\end{equation}
Substituting \eqref{eq:exp_1} in \eqref{eq:EL_eq_2} and solving for the coefficients we get the solution to be,
\begin{equation}
       \theta(r)=\theta_0-\frac{\cot (\theta_0)}{2 r^2}+\frac{\theta_3}{r^3}-\frac{\cos (2 \theta_0) \cot (\theta_0) \csc ^2(\theta_0)}{8 r^4}+ \dots \label{thetaseriesAdS4}
\end{equation}
Finally Substituting this into the Lagrangian in the area functional \eqref{eq:area_func_2}, we get the following expansion
\begin{equation}\label{eq:lag_1}
     \mathcal{L}=\; \sin (\theta_0)-\frac{\sin (\theta_0)}{2 r^2}-\frac{2 (\theta_3 \cos (\theta_0))}{r^3}+\dots 
\end{equation}
Note that we have two integration constants in the final result, $\theta_0$ and $\theta_3$. 

One can also choose to work in $r(\theta)$ language, by expanding around the deepest approach location in the bulk, $\theta=0$. These calculations have been worked out in Appendix \ref{App:RT}. It turns out that in this language, the system can be solved in closed form and the result is presented in eqn. \eqref{r(theta)}. In fact, \eqref{r(theta)} can be inverted to show that it is identical to \eqref{eq:ads_3_sol} -- as mentioned in the previous subsection, the RT surface associated to a spherical subregion can be written in a form independent of dimension. 

The closed form expressions \eqref{r(theta)} or \eqref{eq:ads_3_sol} allow us to compare the solution against \eqref{thetaseriesAdS4} by expanding $\theta$ at large $r$. It turns out that the result is precisely \eqref{thetaseriesAdS4} once we set $\theta_3=0$. The mechanism here is in fact what we outlined at the end of the previous subsection -- the expansion near $r \rightarrow \infty$ is not automatically aware of the conditions one needs to impose in the bulk, and this leads to an extra integration constant. 
A similar phenomenon occurs in flat space as well, and in empty flat space we will check this more concretely -- we can exploit the fact that the minimal surfaces associated to symmetric spi-subregions are hyperplanes and not catenoids (or their higher dimensional generalizations).

Let us make a couple of comments. First of all, in the limit $r\rightarrow\infty$, $\theta(r)=\theta_0$ which is a free parameter. It is in fact the same quantity $\theta_\infty$ discussed in the previous sections. We will later see that the dependence in the flat space calculations is different. The RT data is described differently by a spi-subregion. Secondly, the entropy which is proportional to $\mathcal{A}$ scales as $r^{d-2}$ in AdS$_{d+1}$, i.e. the entropy grows as an `area' on the cut-off surface -- it is a co-dimension 3 quantity. This result is expected for a CFT, when we interpret it as a UV cut-off. Later we will see that unlike here, for flat space calculations the entropy grows as the volume of the  cut-off surface, a co-dimension 2 quantity.

\subsection{Black Hole in the Bulk}

\begin{figure}[h]
    \centering
    \includegraphics[width=9cm]{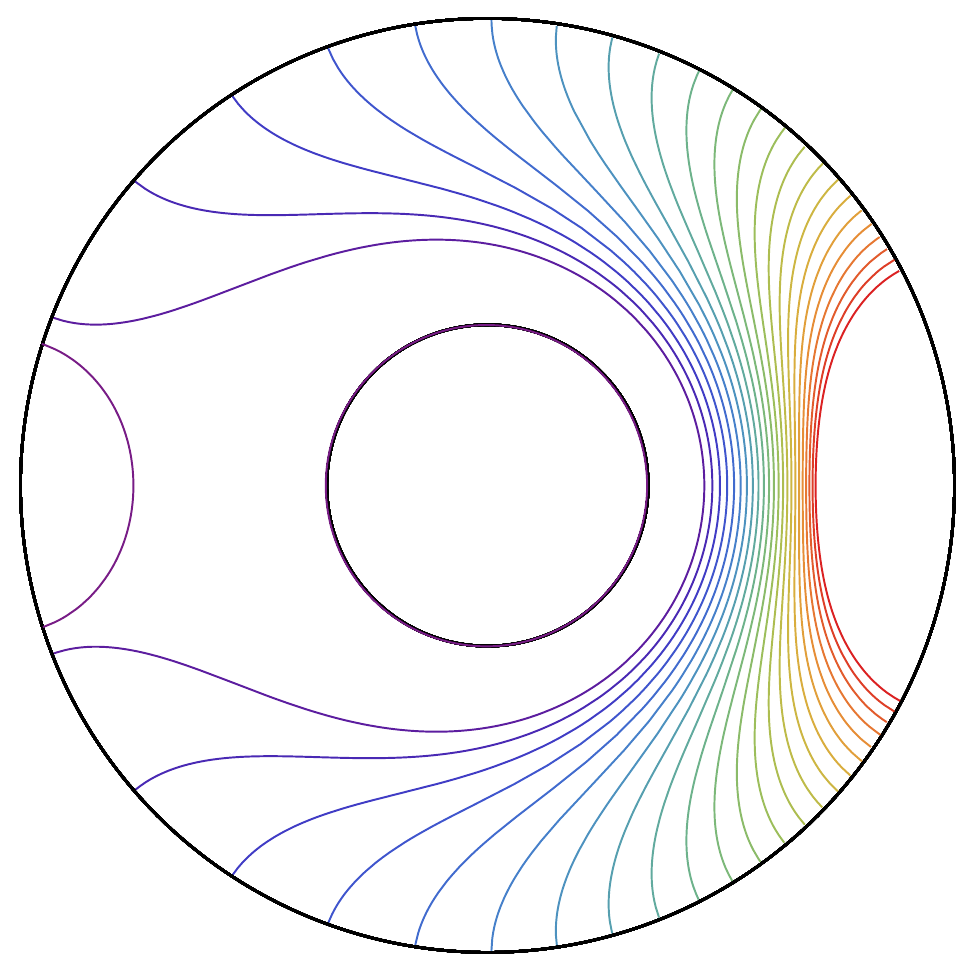}
    \caption{A Plot of RT surfaces for different subregion sizes. The plot is in conformal coordinates. As the subregion size increases the surface penetrates deeper into the bulk. After a critical size the disconnected RT-surface dominates over the connected surface. We have chosen a horizon radius that is small compared to the AdS length to illustrate that the RT surfaces need not probe too close to the horizon.}
    \label{fig:BH_waist}
\end{figure}

The Ryu-Takayanagi prescription can also be used to calculate entanglement entropy associated with a subregion when there is a black hole in the bulk. In this case, the boundary theory is in a mixed state. Therefore the entanglement entropy (and the RT surface) associated to a subregion  and its complement are not the same. This feature is reflected by RT surfaces as shown in figure \ref{fig:BH_waist}. 

It is often the case that subregions have more than one extremal surfaces. In such situations the prescription suggests that the surface with the minimal area is the acceptable RT-surface. One such case is that of a large spherical subregion around an AdS black hole. For large enough subregions the RT-surface is a disjointed surface which is a union of the RT surface associated with the complement subregion and the black hole horizon as in figure \ref{fig:BH_waist}. As, the $\theta_\infty$ increases beyond certain $\theta_\text{critical}$, the dominant curve transitions from the connected to the disconnected surface. This can represented as \cite{Freivogel}
\begin{equation}
    \mathcal{A}(\theta_\text{critical})=\mathcal{A}(\pi-\theta_\text{critical})+\Omega_{d-2}r_h^{d-2}
\end{equation}
where $\mathcal{A}$ is the area functional. As the size of the subregion $A$ increases and becomes the complete boundary region (i.e. in the limit that size of the complement subregion $B\rightarrow0$), the RT-surface is just the black hole horizon and therefore we can conclude that the entropy associated with the thermal state in the boundary theory is proportional to the area of the black hole horizon. 

These are well-known results of the AdS/CFT correspondence when the black hole is large, ie., when the horizon radius satisfies $r_h \gtrsim L$. One fact that will be important for us is that similar statements are in fact true, even for small black holes. There are some differences in detail between the two cases -- eg., the $\theta_{critical}$ happens at a much larger value than $\pi/2$ when $r_h \gg L$, while as $r_h \rightarrow 0$ it gets closer and closer to $\pi/2$. Another distinction is that as the black hole becomes smaller, the RT surface does not probe too close to the horizon. These observations were made in \cite{Freivogel}. One of our results will be that RT-surfaces in asymptotically flat space will show very similar phase transitions in the presence of an IR cut-off $r_{cut}=R_0$, with the role of $r_h/L$ replaced by $r_h/R_0$. The behavior of the phase transitions is qualitatively similar in flt space as well, once we make these replacements. RT surfaces in flat space black holes have not been studied in quantitative detail previously, but see \cite{Chethan_2} for some qualitative comments. 

\subsubsection{BTZ}

Again, our goal is to emphasize the points which have close counterparts or distinctions compared to flat space black holes\footnote{In flat space there are no black holes in 2+1 dimensions, but there are black holes in dimensions starting with 3+1.}. Just as in the empty AdS$_3$ case, RT-surface for BTZ geometry can be calculated analytically. The metric is,
\begin{equation}
    ds^2=-\left(\frac{r^2-r_h^2}{L^2}\right)dt^2+\left(\frac{r^2-r_h^2}{L^2}\right)^{-1}dr^2+r^2d\theta^2
\end{equation}
with similar symmetry assumptions as in section \eqref{AdS_3}, the area functional to be extremized is,
\begin{equation}\label{eq:area_func_3}
    \mathcal{A}=\int\sqrt{\left(\frac{r^2-r_h^2}{L^2}\right)^{-1}r'^2+r^2}\;d\theta=\int\sqrt{\left(\frac{r^2-r_h^2}{L^2}\right)^{-1}+r^2\theta'^2}\;dr
\end{equation}
Since $\theta$ is the cyclic coordinate we get the Euler-Lagrange equations as 
\begin{equation}\label{eq:EL_eq_3}
    \frac{d\theta}{dr}=\frac{cL}{r\sqrt{(r^2-c^2)(r^2-r_h^2)}}
\end{equation}
Integrating \eqref{eq:EL_eq_3} and demanding $r=r_*$ for $\theta=0$ sets $c=r_*$. The final equation of the curve is,
\begin{equation}
    \theta =\frac{L}{r_h}\tan^{-1}{\left(\frac{r_h}{r_*}\sqrt{\frac{r^2-r_*^2}{r^2-r_h^2}}\right)} \label{BTZsol}
\end{equation}
where, $r_*$ is the closest distance from from the origin and is related to $\theta_\infty\;$ as $\;\theta_\infty=L/r_h\tanh^{-1}(r_h/r_*)$ \cite{Freivogel}. As before the integration constant we get in \eqref{eq:EL_eq_3} is set to zero as it just rotates the curve about the origin. Substituting this solution into the area functional \eqref{eq:area_func_3} we get the entropy as 
\begin{align}
    \mathcal{S}=\frac{\mathcal{A}}{4G}&= \frac{1}{4G}\; 2L \ln{\left(\sqrt{r^2-r_*^2}+\sqrt{r^2-r_h^2}\right)}\Bigg|_{r_*} ^{1/\epsilon}\nonumber\\
    &= \frac{L}{2G}\ln{\left(\frac{2}{\epsilon\sqrt{r_*^2-r_h^2}}\right)}=\frac{c}{3}\ln{\left(\frac{2}{\epsilon r_h}\sinh{\left(\frac{r_h}{L}\theta_\infty\right)}\right)}
\end{align}
where $1/\epsilon$ is large distance radial cut-off. After the coordinate change $r=L \sinh{\rho}$ one can see, the final result is in agreement with standard CFT$_2$ results and the cut-off gets interpreted as a UV cut-off in the CFT \cite{RT1, RT2}. Some of the comments we made in empty AdS have counterparts here, which we will not repeat. But one statement does {\em not} have a counterpart -- it is no longer true that the solution \eqref{BTZsol} generalizes to (black holes in) higher dimensions. 

\subsubsection{AdS$_4$ Schwarzchild}

Similar to section \eqref{sec:ads_4}, we can calculate RT-surfaces in higher dimensions as a power series expansion. The AdS$_{d+1}$ black hole metric is,
\begin{equation}
    ds^2=-f_d(r)dt^2+\frac{1}{f_d(r)}dr^2+r^2 d\Omega_{d-1}^2 \label{AdSBH} 
\end{equation}
where the blackening factor is,
\begin{equation}
    f_d(r)=r^2+1-\frac{r_s^{d-2}}{r^{d-2}}  \left(r_s^2+1\right) \label{blackening}
\end{equation}
For $d = 3$, the area functional for the spherical subregion is,
\begin{equation}\label{eq:area_func_4}
    \mathcal{A}=2\pi \int \underbrace{r(\theta) \sin(\theta) \sqrt{\frac{r'(\theta)}{f_4(r)}+r^2}}_\mathcal{L} \; d\theta
\end{equation}
From \eqref{eq:area_func_4} we can write the Euler-Lagrange equation as,
\begin{equation}\label{eq:EL_eq_4}
    \frac{\partial }{\partial r}\left(r \sin(\theta (r)) \sqrt{\frac{r'(\theta)}{f_4(r)}+r^2}\right)=\frac{d}{d \theta}\frac{\partial }{\partial r'}\left(r \sin(\theta (r)) \sqrt{\frac{r'(\theta)}{f_4(r)}+r^2}\right)
\end{equation}
With a similar method as in sec \eqref{sec:ads_4}, we get the solution to be,
\begin{equation}
    \theta(r)=\theta_0-\frac{\cot (\theta_0)}{2 r^2}+\frac{\theta_3}{r^3}+\frac{\cot (\theta_0)-\cot ^3(\theta_0)}{8 r^4}-\frac{\frac{1}{4} \cot (\theta_0) \left(h^3+h-22 \theta_3 \cot (\theta_0)\right)+2 \theta_3}{5 r^5}+ \dots
\end{equation}
On substituting this $\theta(r)$ back into the Lagrangian of \eqref{eq:area_func_4} we get,
\begin{equation}\label{eq:lag_2}
    \mathcal{L}=\; \sin (\theta_0)-\frac{\sin (\theta_0)}{2 r^2}+\frac{h^3 \sin (\theta_0)+h \sin (\theta_0)-4 \theta_3 \cos (\theta_0)}{2 r^3}+\dots
\end{equation}

Many of the features we emphasized for the RT surface solution in empty AdS case are also present here. We note one crucial point: the divergent terms of \eqref{eq:lag_2} are independent of the black hole radius. Not only that, the divergent terms are exactly same as in the empty AdS case. This is an important point and will later turn out to be a crucial difference between the AdS and the flat space RT-surfaces. The fact that divergences are not sensitive to the state in AdS, is a holographic hint that the dual is a local quantum field theory. We will see that this is {\em not} the case in flat space -- the divergences will indeed depend on the state. This is yet another suggestion that the hologram of flat space is non-local.

\section{Ryu-Takayanagi Data in Flat Space}

Holography in flat space is similar enough to AdS that it {\em may} be tractable, but it has enough fundamental differences from AdS that the task is certainly not trivial. One observation made in \cite{Chethan_1, Chethan_2} was that even in asymptotically flat space, one can construct an analogue of an AdS-causal wedge, called the ACD. In this section we will first discuss the conformal compactification of Minkowski space which will help us better compare and contrast the properties of ACDs and AdS-causal wedges and define the notion of a spi-subregion.  We will be careful to keep track of some scales which are often set to unity in these discussions.

\subsection{Conformal Structure of Flat Space}
Let us first set up the conformal structure of flat space with which we will work. We largely follow the conventions of \cite{Hawking_Ellis} while being more careful about dimensions and also working in general dimensions. We start with the flat space Mink$_{d+1}$  metric in $d+1$ dimensions,
\begin{equation}\label{eq:4.1}
    \mathrm{d}s^2=-\mathrm{d}t^2+\mathrm{d}r^2+r^2\mathrm{d}\Omega_{d-1}^2
\end{equation}
where $\mathrm{d}\Omega_{d-1}^2$ is the metric on $\mathbb{S}_{d-1}$. For compactification we use the following coordinate transformations,
\begin{align}
    t+r&=\Lambda\tan\left(\frac{t'+r'}{2}\right)\nonumber\\
    t-r&=\Lambda\tan\left(\frac{t'-r'}{2}\right)
\end{align}
where, $\Lambda$ is a dimensional parameter. This parameter is generally ignored in textbook discussion but is important for our discussion as it sets the scale. This scale we will sometimes relate to the radius of a screen/cut-off. The range of the new coordinates are
\begin{equation}
    -\pi<t'+r'<\pi,\;\quad -\pi<t'-r'<\pi,\quad \; r'>0
\end{equation}
\begin{figure}[h!]
    \centering
    \includegraphics[width=8cm]{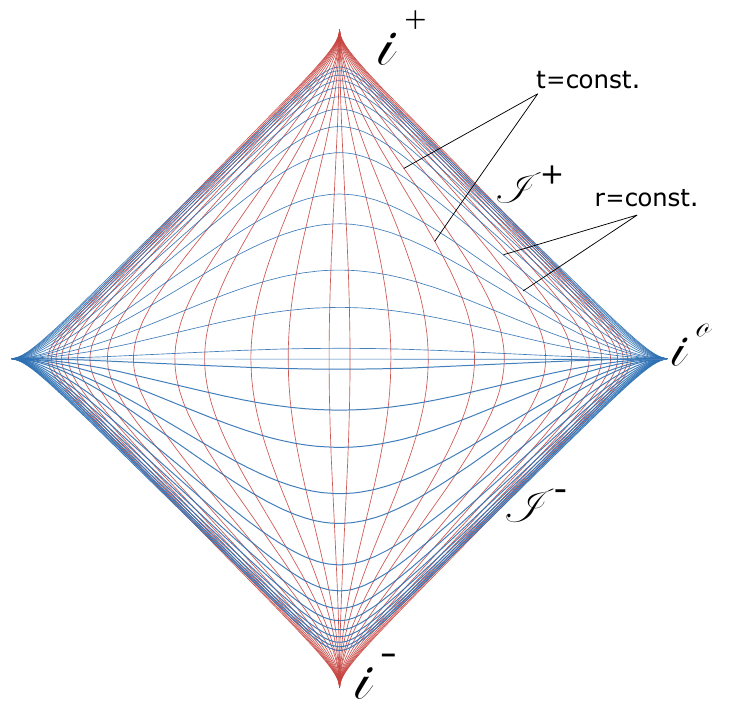}
    \caption{ A plot of the $(t',r')$ coordinates of 2-dimensional flat space. The blue lines are the constant time slices and the red line are constant space slices.}
    \label{fig:penrose_diagram}
\end{figure}
The metric becomes,
\begin{equation}\label{eq:conf_met}
    \mathrm{d}s^2=\Omega^2(-\mathrm{d}t'^2+\mathrm{d}r'^2+\sin^2{r'}\,\mathrm{d}\Omega^2)
\end{equation}
where, $\Omega$ is the conformal factor
\begin{equation}
    \Omega=\frac{\Lambda^2}{4}\sec^2\left(\frac{t+r}{2}\right)\sec^2\left(\frac{t-r}{2}\right) \label{conf-factor}
\end{equation}
 The relation between these coordinates can be given as,
\begin{subequations}\label{eq:conf_mink}
    \begin{align}
        \tan{t'}&=\frac{2\Lambda t}{\Lambda^2-t^2+r^2}\label{eq:4.50}\\
        \tan{r'}&=\frac{2\Lambda r}{\Lambda^2-r^2+t^2}\label{eq:4.60}
    \end{align}
\end{subequations}
These are noteworthy formulas. On comparing \eqref{eq:conf_mink} with \eqref{eq:bound_mink} we see that if we consider the flat space at the boundary of AdS, this dimension $\Lambda$ is automatically the IR scale $L$ of AdS. This observation is interesting because it is another suggestion that $\Lambda$ is best interpreted as an IR scale. 

Later in this section we will set $\Lambda$ to be the radius of the cut-off which is the IR scale of our problem. For further reference we also define $p$ and $q$ via
\begin{equation}
    t' = p+q \quad\quad\quad r'=p-q
\end{equation}
These coordinates will be useful for working with ACDs.

\subsection{Asymptotic Causal Diamonds}

We want to construct analogues of AdS-causal wedges for asymptotically flat space. This can be done using the idea of an Asymptotic Causal Diamond (ACD) introduced in \cite{Chethan_1, Chethan_2}.

Let us start by clarifying the distinction between an AdS-causal wedge and an AdS-Rindler wedge in our nomenclature. We will use the term AdS-causal wedge for bulk causal wedges anchored to the boundary of general {\em asymptotically} AdS spacetimes. When we restrict our attention to empty AdS, they will reduce to AdS-Rindler wedges\footnote{AdS isometries can be used to map any AdS-Rindler wedge to another. Some related facts are discussed in Appendix \ref{AppA}.}. Often we will be interested in boundary subregions which are spherical. The boundary projections of AdS-causal wedges in such cases will simply be boundary causal diamonds associated to these spherical subregions. These can equivalently be defined via two points on the boundary -- the future and past vertices of this boundary causal diamond. One of the main observations of \cite{Chethan_1} was that even though there is no causal structure on the null boundary of flat space, the idea of working with two points (one each on $\scrip$ and $\scrim$) generalizes in a useful way. The analogues of {\em bulk} causal wedges can be defined using these two vertices even in asymptotically flat space and these are what ACDs are. 

We will often be interested in symmetric ACDs. These are obtained by choosing the vertices symmetrically on $\scrip$ and $\scrim$ around a convenient $t=0$ slice in the bulk\footnote{This (useful) choice of the $t=0$ slice can depend on the physics we are considering -- if we wait long enough, black holes can evaporate away, for example.}. They can be defined via
\begin{equation}
    \mathfrak{C}=\mathfrak{I}^-(Q) \cap \mathfrak{I}^+(-Q), \; {\rm where} \; Q \in \scrip, \ -Q \in  \scrim
\end{equation}
where $\mathfrak{I}^+$ and $\mathfrak{I}^-$ denote future and past light cones of a point in the spacetime manifold, extended to include the conformal boundaries. We do not have to work necessarily with symmetric ACDs  -- we can choose the past and future points to be $P$ and $-Q$ instead of $Q$ and $-Q$, say -- but it is often sufficient to do so. In particular in empty Minkowski space this corresponds to a choice of time slice which can be reached via an isometry. 

\subsubsection{Rindler Wedge as an ACD in Empty Minkowski Space}

Just as the AdS-causal wedge becomes more tractable in the case of empty AdS where it reduces to AdS-Rindler, the ACD becomes simpler in Minkowski space. 

It is useful to start with a bulk construction to get some intuition about ACD. We start with a causal diamond in the bulk defined as the intersection of the past light cone of a point $p_F$ with time coordinate $t > 0$ and the future light cone of a point $p_P$ with $t < 0$. This is a bulk causal diamond and it can be shown to be the causal development of a spherical region \cite{Chethan_1}
\begin{equation}
    \mathcal{C}\equiv \mathfrak{I}^{+}(p_P)\cap \mathfrak{I}^{-}(p_F)
\end{equation}
We will call the points $p_P$ and $p_F$ the ``vertices" of the causal diamond. When the points are placed symmetrically around the $t=0$ slice, it is a symmetric causal diamond. The location of the vertices of a symmetric causal diamond are of the form,
\begin{equation}
    p_F=(t,\,x_1,\,x_2, \, ...\, ,\, x_{d}) \quad\quad\quad p_P=(-t,\,x_1,\,x_2, \, ...\, , \,x_{d})
\end{equation}
in $d+1$ dimensions. The (past) light cone of a point $(T,R,\theta',\phi_1',\, ... \, ,\phi_{d-2})$ is given by,
\begin{equation} \label{eq:bulk_cd}
   -(t-T)^2 + r^2+R^2-2 r R \Bigl(\cos \theta \cos \theta'+\sin \theta \sin \theta' \bigl(\; ...\;\bigr)\Bigr)=0
\end{equation}
where the $...$ can be explicitly written down but are unnecessary for our present purposes. 
Without loss of generality we can always choose our axes such that $p_F$ aligns with the pole. This sets $\theta'=0$ and we get,
\begin{equation}\label{eq:cone}
   -(t-T)^2 + r^2+R^2-2 r R \cos (\theta)=0
\end{equation}
In conformal coordinates we have the following relations,
\begin{align}
    t&=\Lambda\frac{\tan{p}+\tan{q}}{2}&r=\Lambda\frac{\tan{p}-\tan{q}}{2}\nonumber\\
    T&=\Lambda\frac{\tan{P}+\tan{Q}}{2}&R=\Lambda\frac{\tan{P}-\tan{Q}}{2}\nonumber
\end{align}

Now that we have the equation for the surface of  this causal diamond, we will send the vertex to $\scrip$ keeping $Q$ fixed. This we can do by taking the limit $P\rightarrow \pi/2$. Writing \eqref{eq:cone} in conformal coordinates and taking the limit we get, 
\begin{equation}\label{eq:acd_conf}
    (\tan{p}+\tan{q})-(\tan{p}-\tan{q})\cos{(\theta)}-2\tan{Q}=0
\end{equation}
Or in Minkowski coordinates,
\begin{equation}\label{eq:acd}
    t-r\cos{(\theta)}-\Lambda\tan{Q}=0
\end{equation}
the coordinate $Q$ decides the location of the vertex on $\scrip$. Thus, we finally end up with a causal wedge attached to the $\scrip$, which is the ACD.  Note that \eqref{eq:acd} is the equation only for the past light cone of the ACD. We can obtain the equation of the future light cone by setting $(t \rightarrow -t)$ (note that the past vertex of the symmetric ACD incorporates $Q \rightarrow -Q$ so there is no need to do that in the equation).

\begin{figure}[h!]
    \centering
    \includegraphics[width=7cm]{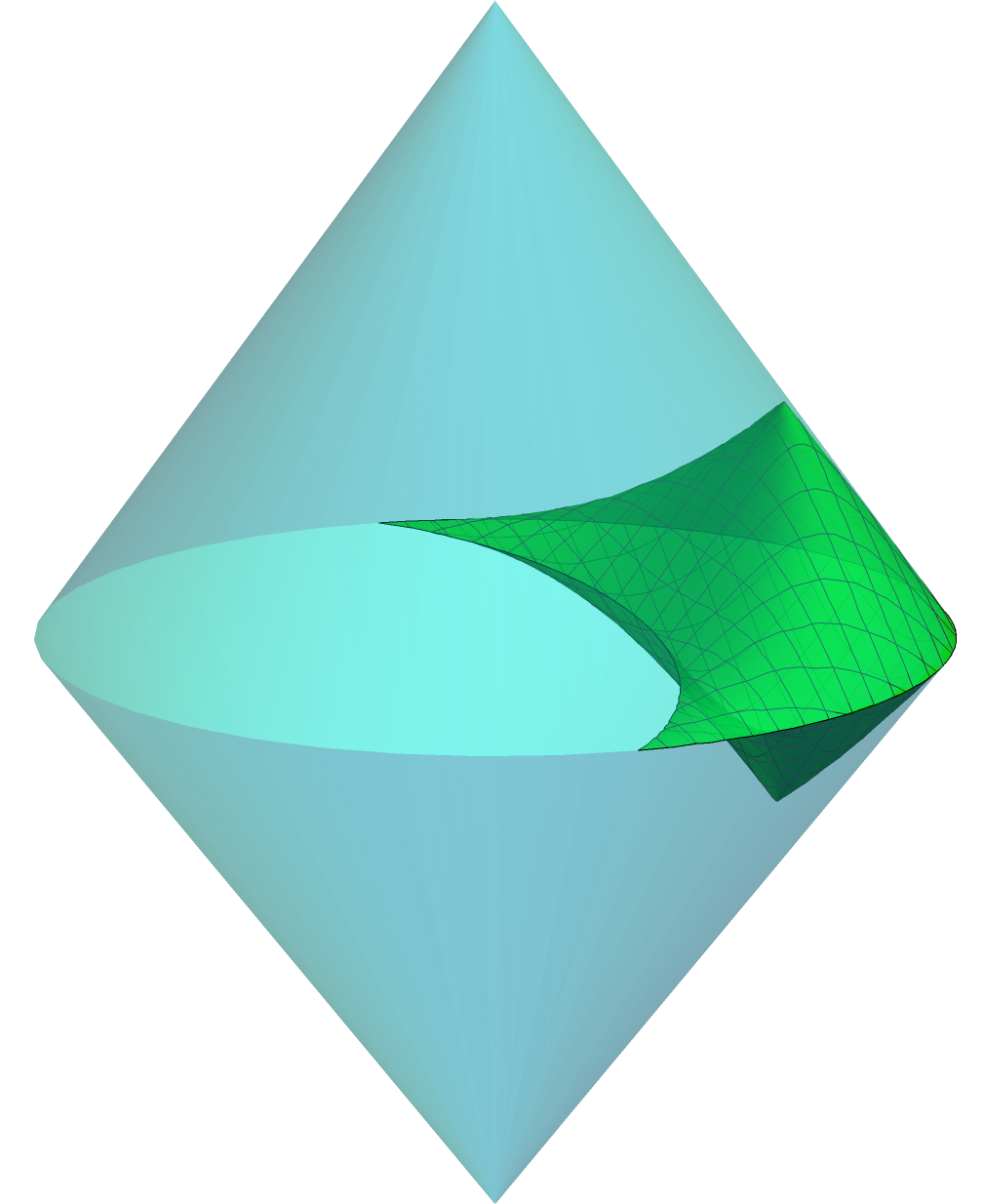}
    \caption{The plot of a symmetric ACD (green) inside the conformal diagram of empty Minkowski space.}
    \label{fig:acd}
\end{figure}

An ACD in Minkowski space is (up to isometries) a Rindler wedge of flat space, but with crucial emphasis on the boundary behavior of the wedge rather than the bulk behavior (as is often the case in usual discussions). The connection with Rindler wedge can be seen from \eqref{eq:acd} that an ACD is the causal development of the spatial region $r \cos{\theta} > \Lambda \tan{Q}$. It is instructive contrasting the structure of an ACD with that of the AdS-Rindler wedge. The intersection of an ACD with the boundary is a co-dimension 2 surface unlike in the AdS-Rindler wedge case where its intersection with the AdS boundary is a co-dimension 1 surface. This can be seen by setting the limit $p\rightarrow\pi/2$ in \eqref{eq:acd_conf} after solving for $\cos \theta$. We end up with 
\begin{equation}
    \cos{(\theta)}=1
\end{equation}

One of the most fundamental differences between a Minkowski ACD and an AdS-Rindler wedge is in their waist. The waist of the symmetric ACD can be simply obtained by setting $t=0$ in \eqref{eq:acd}
\begin{equation}\label{eq:waist_1}
    r\cos{(\theta)}=-\Lambda\tan{Q}
\end{equation}
This is just a hyperplane -- unsurprising, because in empty flat space the causal surface is also the extremal surface and therefore has minimal area. 
\begin{figure}[h]
\centering
     \begin{subfigure}[t]{0.45\textwidth}
            \centering
            \includegraphics[width=7.5cm]{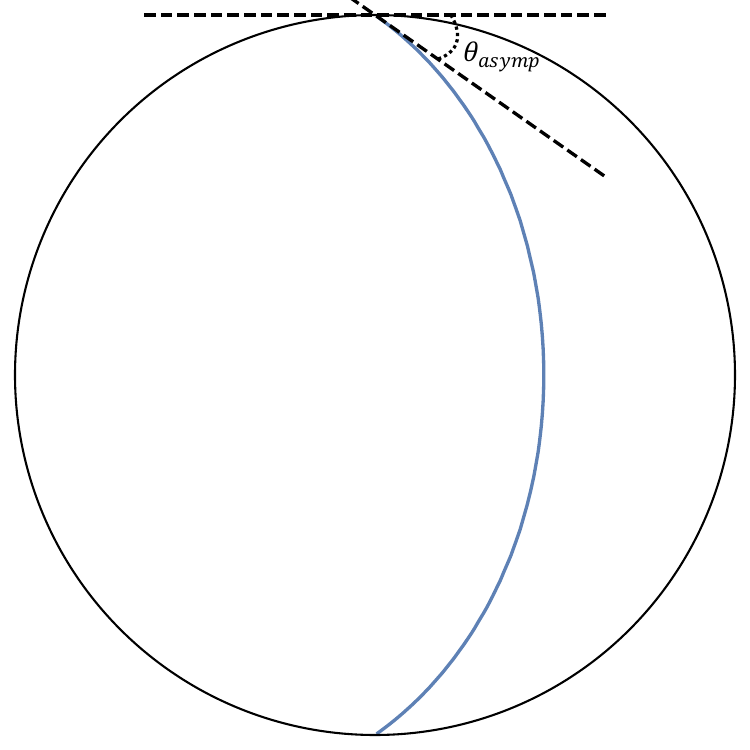}
            \caption{}
            \label{fig:empty_flat_2}
     \end{subfigure}
     \hfill
      \begin{subfigure}[t]{0.5\textwidth}
            \centering
            \includegraphics[width=7.5cm]{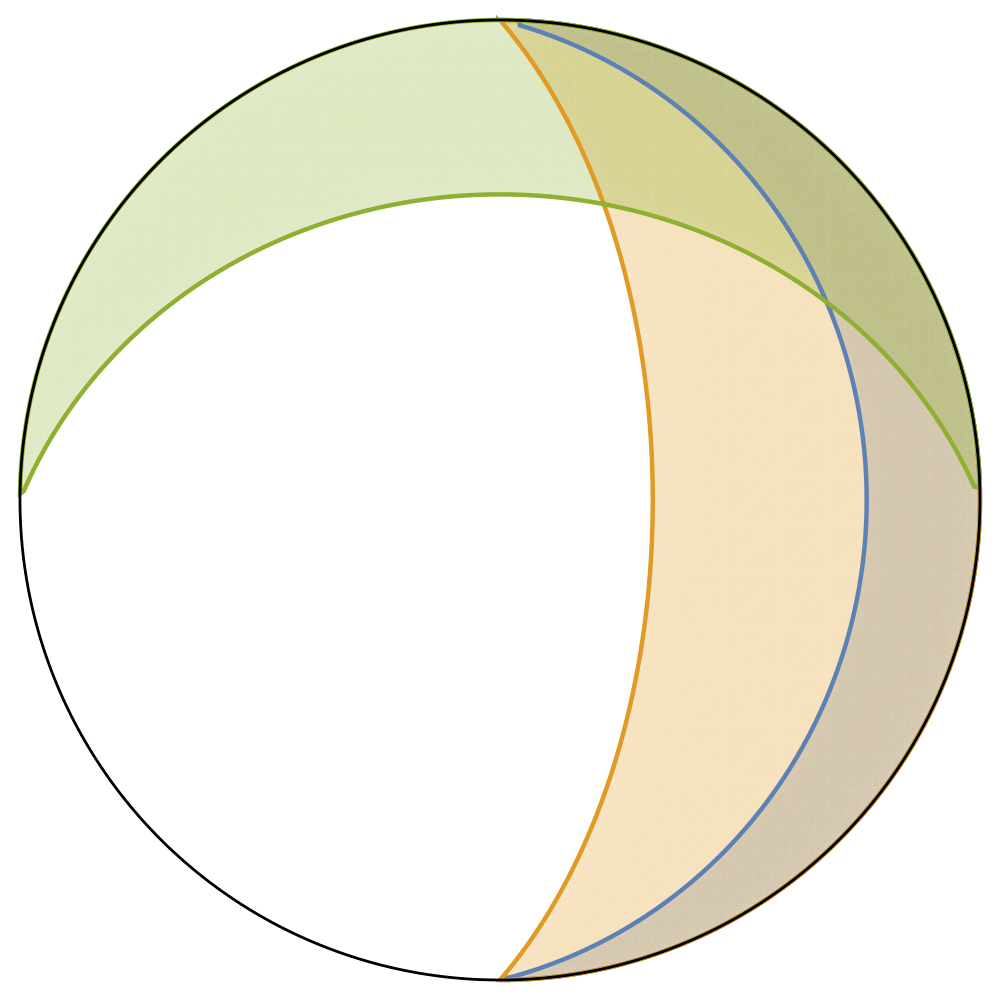}
            \caption{}
            \label{fig:empty_flat}
     \end{subfigure}  
     \caption{(a) The blue curve is the waist of an ACD in Minkowki space. It is also the RT surface associated to the boundary data $\theta_{asymp}$, and is plotted by treating $r'$ and $\theta$ as polar coordinates in \eqref{eq:waist_conf}. The data specified by $ \theta_{asymp}$ is equivalent to that in $\tan Q$. But the former can be viewed as data intrinsic to spi, while the latter is more naturally thought of as data on $\scri$. (b) The waists of the ACD have been plotted for different values of the vertex location. As the vertex moves towards future timelike infinity, the waist becomes wider and eventually the entire spacetime ends up in the causal diamond.}
     \label{fig:acd_waist}
\end{figure}
In conformal coordinates this surface is given by \footnote{Note that $t'=0$ and $t=0$ coincide in the two coordinates.},
\begin{equation}\label{eq:waist_conf}
    \tan{(r'/2)}\cos{(\theta)}=-\tan{Q}
\end{equation}
Unlike in AdS the waist always lands on antipodal points on spi in the conformal diagram. We will discuss this point in detail in the next sub-section. 

\begin{figure}[h]
    \centering
    \begin{subfigure}[t]{0.45\textwidth}
        \centering
        \includegraphics[width=6cm]{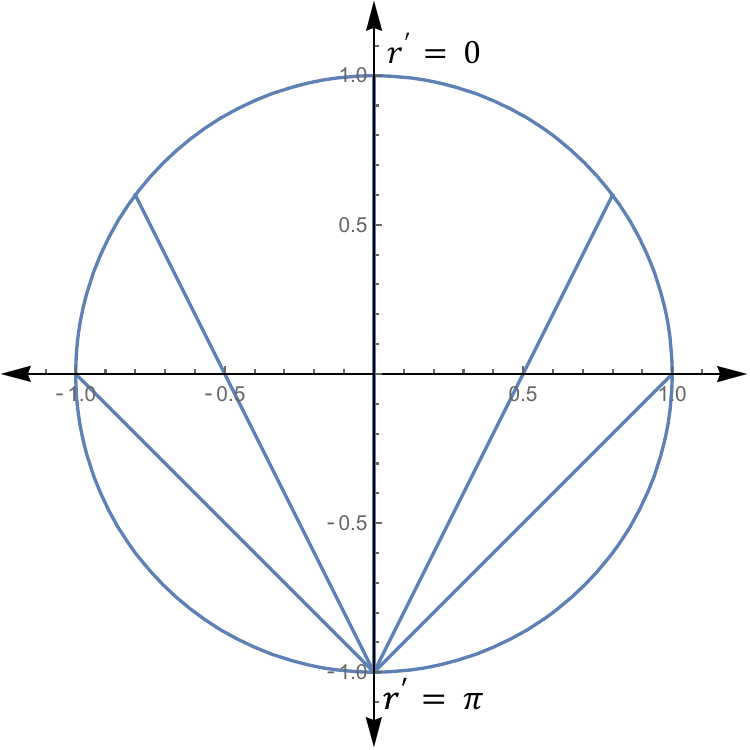}
        \caption{}
    \end{subfigure}
    \begin{subfigure}[t]{0.45\textwidth}
        \centering
        \includegraphics[width=6cm]{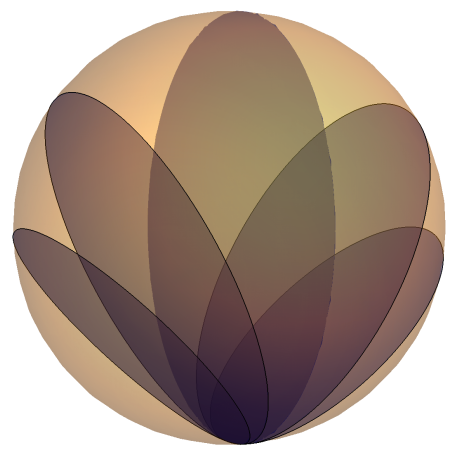}
        \caption{}
    \end{subfigure}
    \caption{(a) The $t=0$ slice of compactified Minkowski space in $1+1D$. The poles are at $r'=0$ (North) and $\pi$ (South). South pole is the point at infinity of $r$. (b) The $t=0$ slice of compactified Minkowski space in $2+1D$. The South pole is the  point at infinity of $r$ in higher dimensions also.}
    \label{fig:re_cast}
\end{figure}

The coordinate $r'$ is often treated as an angle since \eqref{eq:conf_met} is the metric of a cylinder with $r'$ as the polar angle. With this interpretation in mind we can re-cast \eqref{eq:waist_conf} as,
\begin{equation}\label{eq:re_cast}
    1+z=-\frac{x}{\tan(Q)}.
\end{equation}
This equation interprets the waist of an ACD as the intersection of the unit sphere $x^2+y^2+z^2=1$ with the plane given by \eqref{eq:re_cast}\footnote{This discussion is specific to 2+1 dimensional Minkowski space, but the discussion generalizes to $N$ dimensions trivially upon replacing $z$ with the Cartesian direction to the North pole, $x_N$, and replacing $x$ with $\sqrt{x_1^2+x_2^2+...+x_{N-1}^2}$ in \eqref{eq:re_cast}. We will often discuss things in 3+1 dimensional Minkowski space where the subtleties we care about are already visible.}. This sphere is essentially the $t=0$ spatial slice and the south pole ($r'=\pi$) is the point at infinity of $r$. This $r \rightarrow \infty$ is a ``point" in all dimensions as illustrated in the Fig \ref{fig:re_cast}.

These discussions are in the setting of empty Minkowski space, but the asymptotic data defined via conformal coordinates can be used as holographic data more generally in asymptotically flat spacetimes. This is done via the notion of a spi-subregion.

\subsection{Symmetric Spi-Subregions}\label{spisub}

Due to the absence of a well-defined boundary causal structure, identifying analogues of boundary subregions is not straightforward in flat space. So we first observe that in an asymptotically AdS spacetime, subregions on the boundary can also be characterized via the minimal surfaces anchored to them. These minimal surfaces are the waists of AdS-causal wedges which can also be viewed (in empty AdS) as the horizons of AdS-Rindler. This suggests that we should use the waist of an ACD in empty Minkowski space as a tool for characterizing ``subregion data" in flat space. We will first do this for the flat space analogue of a spherical subregion -- we will call the resulting object, a symmetric spi-subregion. In the next subsection we will generalize the discussion to general spi-subregions. It turns out that waists of ACDs approach antipodal points on spi at the conformal boundary. For symmetric spi-subregions, these waists tend to become (hyper)planar at infinity\footnote{Note that there are non-trivial minimal surfaces in $d >2$. What happens to these minimal surfaces in the asymptotic limit, is related to the notion of a general spi-subregion. In Mink$_{3}$, minimal surfaces are always straight lines.}. 


Once you fix the origin and the radius of the spherical subregion, the AdS-Rindler wedge is uniquely defined. In other words the position of the vertices of an AdS-Rindler wedge uniquely fixes the boundary subregion. The analogous statement in asymptotically flat space is that a symmetric spi-subregion can be defined via the vertices of a symmetric ACD.  As long as the angular locations of the vertices are left unchanged, changing the null ($Q$ and $-Q$) coordinate of the vertex of a  symmetric ACD on $\scri$ simply changes the angle of approach of the minimal surface at spi in conformal coordinates. See Figure \ref{fig:empty_flat}. Not in particular that the waist always goes to antipodal points. Even though structurally quite different, this angle has some similarities to the $\theta_\infty$ angle we discussed when discussing spherical subregions in AdS. Just as for a spherical subregion in AdS, apart from its ``center'', there is only one real number worth of data that is needed to specify a symmetric spi-subregion. In AdS, we called it $\theta_\infty$, here we will call it the asymptotic approach angle, $\theta_{asymp}$. It defines the symmetric spi-subregion, once we choose the North pole on spi. This $\theta_{asymp}$ is essentially the same data as $\tan Q$, but it is convenient to give it a geometric interpretation intrinsic to spi that will clarify why we call it the approach angle. The data $Q$ on the other hand, is more manifestly associated to the vertices of the ACD and is best viewed as living on $\scri$. If the spacetime differs from empty Minkowski space in the bulk (eg., when there is a black hole) then this $\theta_{asymp}$ is the true holographic data that one should work with in defining an RT surface. We will explicitly do such calculations in the next section. 

Fig \ref{fig:empty_flat_2} gives a geometric definition of $\theta_{asymp}$. It is simply a convenient angle one can define for lines (or hyperplanes) that reach infinity in conformal coordinates. Let us determine it explicitly as a small exercise, and also to connect it to the $Q$ data. We are interested in determining the ``slope'' of the equation for the waist \eqref{eq:waist_conf}, treated as an equation in polar coordinates, at spi. From the geometry of Fig \ref{fig:empty_flat_2} we see that the slope of the curve is
\begin{equation}
    \frac{dy'}{dx'}=\frac{dy'/d\theta}{dx'/d\theta}=\frac{(dr'/d\theta) \sin\theta + r' \cos{\theta}}{(dr'/d\theta) \cos\theta - r' \sin{\theta}}
\end{equation}
From the equation of the waist, we have
\begin{equation}
    \frac{dr'}{d\theta}=-\frac{2 \tan Q \sin{\theta}}{\tan^2 Q+\cos^2{\theta}}
\end{equation}
The slope at $\theta=\pi/2$ is therefore
\begin{equation}
    \text{slope}=\frac{dy'}{dx'}\Big|_{\theta=\pi/2,\; r'=\pi}=\frac{2}{\pi \tan Q}
\end{equation}
\begin{equation}
    \theta_{asymp}=\pi - \tan^{-1}{\left(\frac{2}{\pi \tan Q}\right)}
\end{equation}
The above result is valid when $Q \in \left(0, \frac{\pi}{2}\right)$. After a similar calculation valid for higher vertices, we get the final result
\begin{equation}
\theta_{asymp}=
    \begin{cases}
        -\tan^{-1}\left(\frac{2}{\pi \tan{Q}}\right)  & \text{for } Q \in \left(-\frac{\pi}{2}, 0\right)\\
        \pi-\tan^{-1}\left(\frac{2}{\pi \tan{Q}}\right)  & \text{for } Q \in \left(0, \frac{\pi}{2}\right)
    \end{cases} \label{thetaTan}
\end{equation}

Note that in empty Minkowski space, $\theta_{asymp}$ is related to the closest distance of approach of the minimal surface to the origin of the spacetime. By choosing the coordinates suitably (and without loss of generality) we can relate it to the perpendicular distance between the minimal surface and the $r \cos \theta = 0 $ surface. This is a consequence of the fact that $(-\tan Q)$ is related to the closest distance of approach of the minimal surface to the  origin (see \ref{eq:waist_1}). 

But let us make an important comment. The distance of closest approach in the bulk is a dimensionful quantity (note that there is a $\Lambda$ in (\ref{eq:waist_1})) while $\theta_{asymp}$ is dimensionless. The choice of $\Lambda$ seems arbitrary at this stage, and the asymptotic conformal data (like $\theta_{asymp}$) does not depend on it. But there is an eminently natural choice for $\Lambda$. This fact manifests itself, when we use the $\theta_{asymp}$ data to compute the area of the minimal surface. The area diverges due to the infinite volume of flat space, and this can be regulated by putting a large radius cut-off. With a large radial cut off $r=R_0$, the area of the minimal surface (ie., a co-dimension 2-hyperplane in Minkowski space) is given by,
\begin{equation}\label{eq:area_acd_waist}
    \mathcal{A}=\frac{\pi^{\frac{d-1}{2}}}{\Gamma\left(\frac{d+1}{2}\right)}R_0^{d-1}\left(1-\tan^2 Q\right)^{\frac{d-1}{2}}
\end{equation}
Crucially, in writing this result, we have set $\Lambda = R_0$. For any other choice of $\Lambda$ the formula would have two scales and we will not be able to obtain this simple and natural form by extracting an overall factor of $R_0^{d-1}$. 

We see here that the area and therefore the entanglement entropy go as $r^{d-1}$ which gives us a volume law for the entropy on the screen, instead of the conventional area law. This suggests two things. Firstly, it is an indication that the divergence should indeed be viewed as an IR divergence. Secondly, it suggests that the holographic theory is a non-local theory (for local theories the entanglement entropy goes as area law). Both these facts are conceptually significant, and depart from what one sees in AdS/CFT when one introduces a cut-off at super-AdS scales. This result ties in very naturally with the suggestion \cite{Verlinde} that the holographic duality is an IR/IR duality in flat space and sub-AdS scales, and that the degrees of freedom are not of a local QFT  -- instead the dual degrees of freedom are those of long strings. This perspective immediately explains why the interactions are non-local (strings are not pointlike) and why there are no UV divergences (strings are UV-finite).

There is in fact a second piece of evidence that suggests that the divergence noted above should be viewed as an IR divergence -- we briefly remarked on this earlier, and it arises via conventional AdS/CFT. Consider the Poincare patch of AdS. The boundary of the Poincare Patch is Minkowski space and the Penrose diagram of flat space emerges naturally there. The dimensionful $\Lambda$ that shows up there automatically is the AdS length scale, which we know from AdS/CFT is to be viewed as an IR scale. The short calculation that demonstrates this observation is presented in Appendix \ref{AppA}. 

Let us conclude this subsection by emphasizing again that this $\theta_{asymp}$ is conceptually different from the $\theta_\infty$ discussed in Section \ref{sec:bound_sub} in the AdS setting. The latter is a coordinate value and therefore signifies a location on the conformal boundary whereas $\theta_{asymp}$ is not a physical location where an observer can stand. It is more analogous to a direction of approach to infinity, and therefore is naturally well-defined on spi. Despite this, it naturally allows an interpretation as bi-partitioning spi (or more importantly for us, the associated bulk state).  

Most of our observations in this paper only require the notion of the symmetric spi-subregion that we have introduced in this section, but when $d > 2$ it is possible to generalize this notion. This should be compared to the fact that the boundary of AdS$_3$ has only spherical subregions while in higher dimensions subregions can be more general. Even though not strictly necessary for the rest of the paper, this reveals the richness of spi. We expect that this will be significant when trying to construct a holographic theory at spi, and therefore we develop it further in the next subsection.

\subsection{General Spi-Subregions}

In our discussion so far, we have constructed analogues of causal wedges, spherical subregions and RT surfaces -- all familiar from AdS -- in asymptotically flat spaces. A natural question that arises is whether one can extend this construction to encompass more general sub-regions at spatial infinity. To get an understanding of this question it is useful to look at unions of bulk causal diamonds in their ACD limit. 

We will start with 2+1 dimensions where the situation is straightforward -- despite some technical complications in higher dimensions, we will in fact see that the general ideas are valid across dimensions. The 2+1 dimensional case should be clear from Fig \ref{fig:empty_flat}. The union of two ACD waists is given by the unions of the corresponding regions in that figure. It is clear that the corresponding asymptotic data is given by the two angles of approach at infinity -- they capture one end of one of the spi-subregions and the other end of the other. This is what we will refer to as a non-trivial union. Note that if the ``centers'' of the two symmetric spi-subregions are the same, the union is simply the bigger symmetric spi-subregion. ``Bigger" here simply means the one with the bigger $\theta_{asymp}$.  We will call such a union, a trivial union. It is straightforward to construct a general spi-subregion by arbitrary (trivial and non-trivial) unions using these principles.

An interesting question is -- what is the bulk region that is in the causal wedge of the union of two spi-subregions.  The answer to this question in flat space is distinct from that in AdS, and is therefore more interesting. We claim\footnote{This was briefly suggested in \cite{Chethan_2}.} that whenever the union involves a non-trivial union of two ACDs, the causal wedge is the entire spacetime\footnote{This comment applies to Minkowski space -- of course, when there are black holes etc., there may be regions inside horizons which are not accessible to boundary-anchored RT surfaces. Such features exist in AdS as well, so we will not emphasize them.}. There are a couple of ways in which one can convince oneself that this is the case. The first argument is to note that if we had a screen in the bulk, the RT surfaces associated to the unions of the induced screen subregions go to infinity as the screen size is sent to infinity. This is trivial to see in 2+1 dimensions, where the only minimal surfaces are straight lines. But it is true in higher dimensions as well, as we will argue later in this section. A second argument is to consider unions of three spi-subregions such that the center of the third spi-subregion is in the intersection of the first two. In such a configuration the spi-subregion that is the union of  the three is determined by the first two. Now, one can ``fatten'' the waist of the third ACD by increasing its $\theta_{asymp}$ arbitrarily, or equivalently by sliding the vertices towards timelike infinity. Under this operation, its bulk causal wedge can become arbitrarily large and can cover the entire spacetime. This is a suggestion that the union of the first two should contain the entire spacetime. Note that this observation fails in AdS because as one enlarges a boundary causal diamond that is in the intersection of two other boundary causal diamonds, it will eventually become bigger than the union of the first two. In the analogous construction at the  boundary of Minkowski space, this happens only when the entire spacetime is in the causal wedge as well.

Note that the above two arguments are qualitatively distinct. The first one involves a bulk cut-off, but the other can be phrased entirely in terms of the causal wedges of spi-subregion data. We see repeated hints of this kind in this paper, that the bulk IR cut-off is a natural part of the definition of the hologram of flat space -- just as the UV cut-off is an implicit part of the definition of the CFT in AdS/CFT. Let us also note that the causal/entanglement wedge structure of unions here may be natural in a non-local holographic theory. In a theory with all-to-all couplings like in SYK-like models, bi-partitions of the lattice may be more useful than multi-partitions and  their unions. Our entire discussion seems to strongly suggest that bi-partitions are more interesting in the hologram of flat space, perhaps hinting at the two ends of a string.

\begin{figure}[h!]
    \centering
    \includegraphics[width=10cm]{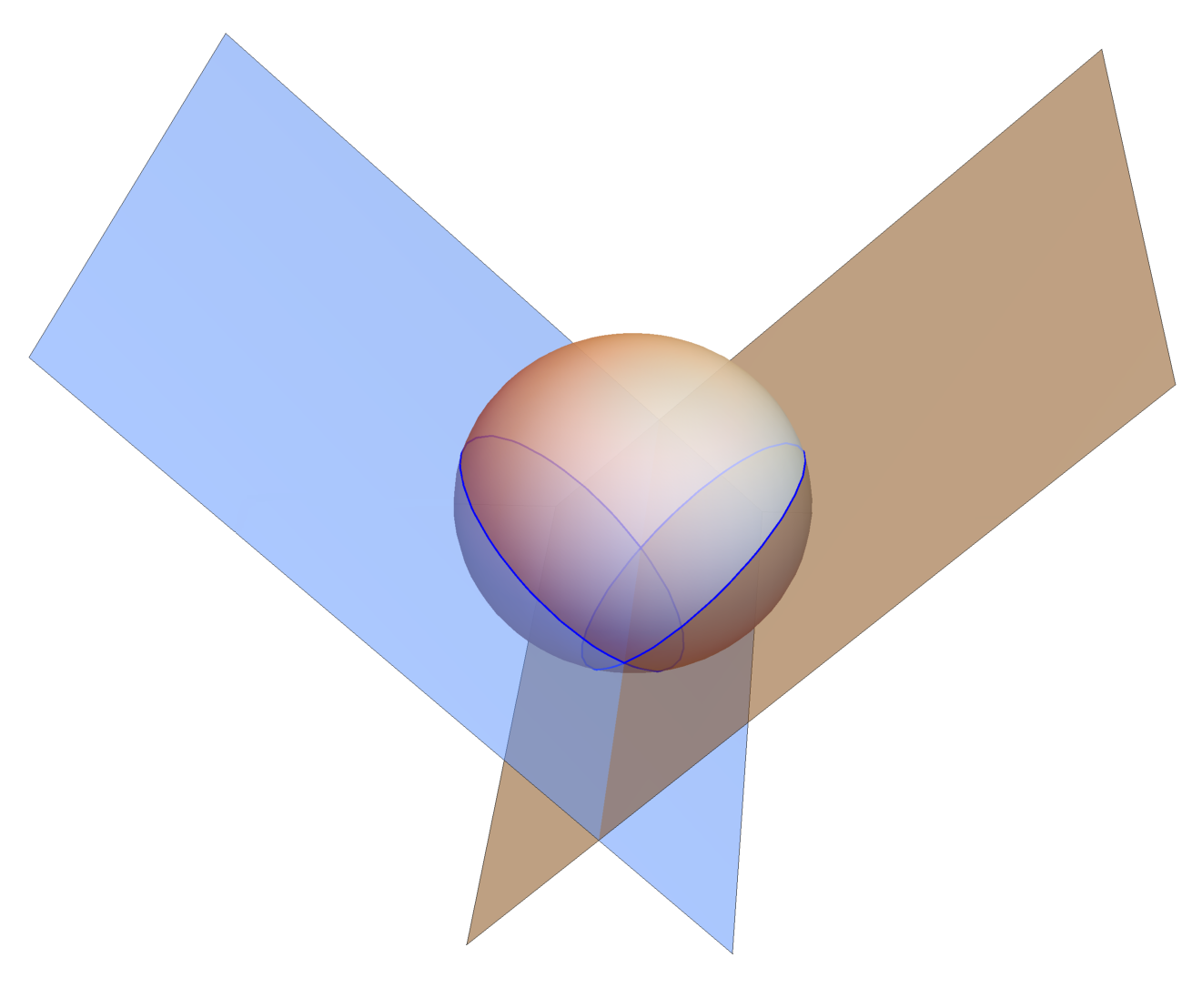}
    \caption{The two planes (blue and brown) are the waists of two ACDs. For a finite radial cut-off, the intersection of the planes and the screen give us a curve on the screen (bold blue curve).}
    \label{fig:acd_union}
\end{figure}

Let us now turn our discussion to higher dimensions. The discussions in all higher dimensions are essentially identical, so we will restrict ourselves to the concrete setting of 3+1 where the subtleties are already visible. Let us consider two bulk causal diamonds with their vertices at arbitrary point in the bulk. Then \eqref{eq:bulk_cd} becomes,
\begin{equation} 
   -(t-t_{1,2})^2 + r^2+r_{1,2}^2-2 r R \Bigl(\cos \theta \cos \theta_{1,2}+\sin \theta \sin \theta_{1,2} \cos{\left(\phi-\phi_{1,2}\right)}\Bigr)=0
\end{equation}
where $(t_1,\,r_1,\,\theta_1,\,\phi_1)$ and $(t_2,\,r_2,\,\theta_2,\,\phi_2)$ are the vertices of the two causal diamonds. Taking the limit of the vertices going to $\scrip$ (keeping the angular coordinates fixed) we get the following equations of the ACDs,
\begin{equation}
    t-r\left(\cos \theta \cos \theta_{1,2}+\sin \theta \sin \theta_{1,2} \cos{\left(\phi-\phi_{1,2}\right)}\right)=\Lambda\tan{Q_{1,2}}
\end{equation}
The resulting waists are two planes, a fact we could have guessed. Generically, they will intersect. With a finite radial cutoff, a curve on the screen emerges from the intersection of these surfaces (see Fig. \ref{fig:acd_union}). A key fact in 3+1 (and higher) dimensions is that there exist non-trivial minimal surfaces with arbitrary curves on spherical screens. This statement is an aspect of the Plateau problem and was proved in extreme generality by Jesse Douglas \cite{Douglas}. In 2+1 dimensions, the minimal surfaces will simply be straight lines and we do not see such a rich structure.  

As the radius of  the spherical cut-off approaches infinity, this curve is the data that captures the information about the (general) spi-subregion. By taking (possibly infinite) non-trivial unions, we can construct arbitrary spi-subregions\footnote{This philosophy is identical to that in AdS, where general subregions can be viewed as (infinite) unions of spherical subregions.}. So we will focus on the case of a single non-trivial union. Let us make it explicit in the context of two ACDs oriented so that their waists (planes) are intersecting at right angle. We will also assume that the closest distance of approach from the origin to either of these planes is $a$, the same constant. This restricted setting is enough to capture the geometric features we wish to illustrate. 

By choosing the coordinates wisely, we can write the two planes as 
\bea
z+x=a, \ \ {\rm and} \ \ z-x=a. 
\eea
In polar coordinates this becomes 
\bea
r(\cos \theta +  \sin \theta \cos \phi) &=& a, \ \ {\rm for}  \ \ \phi \in (0, \pi/2) \cup  (3 \pi/2, 2 \pi) \\
r(\cos \theta -  \sin \theta \cos \phi) &=& a, \ \ {\rm for}  \ \ \phi \in (\pi/2, 3\pi/2). 
\eea
We are no longer working with the North pole aligned with either of the ACDs -- North pole is now a point in the interior of their intersection for convenience. These equations can be translated to conformal coordinates, by simply replacing $r$ with $\Lambda \tan (r'/2)$ and absorbing the scale in $a$ via $a \equiv -\sqrt{2} \Lambda \tan Q$. This leads to a generalization of  \eqref{eq:waist_conf} (see below) and we can read off the $\theta_{asymp}$ from the shape of the curve at $r'=\pi$. This is a conceptually straightforward problem which we solve below for this relatively simple case. The details are a bit ugly -- but the resulting waist structure can be intuitively understood via surfaces of revolutions obtained from Fig. \ref{fig:empty_flat} using say the green and brown regions as examples.

\begin{figure}[h!]
    \centering
    \includegraphics[width=8cm]{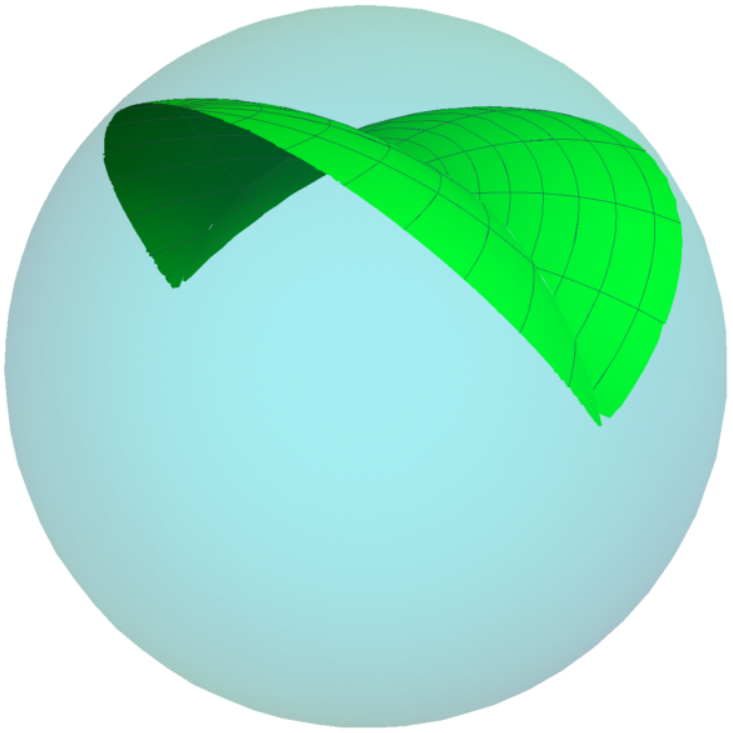}
    \caption{Plot (up to overall rotation) of the surfaces in \eqref{surf1} and \eqref{surf2}, where $r', \theta, \phi$ are treated as spherical polar coordinates.}
    \label{dbl-waist}
\end{figure}

It is easy to see that both pieces of the curve live on planes in the bulk that are a distance $(-\tan Q)$ from the origin of conformal coordinates\footnote{We introduced a $\sqrt{2}$ in the relation relating $a$ and $\tan Q$ above, to arrange this.}, and therefore the magnitude of $\theta_{asymp}$ is precisely that given in \eqref{thetaTan}.  The task is simply to determine the curve $\theta(\phi)$ at $r'=\pi$. We have the equation of surfaces in conformal coordinates,
\begin{align}
    \tan{(r'/2)}(\cos{\theta}+\sin{\theta}\cos{\phi}) &= a, \quad \text{for} \quad \phi \in (0,\pi/2) \cup (3\pi/2,2\pi) \label{surf1} \\
    \tan{(r'/2)}(\cos{\theta}-\sin{\theta}\cos{\phi}) &= a, \quad \text{for} \quad \phi \in (\pi/2,3\pi/2) \label{surf2}
\end{align}
for $r'=\pi$ we have
\begin{align}
    cos{\theta}+\sin{\theta}\cos{\phi} &= 0, \quad \text{for} \quad \phi \in (0,\pi/2) \cup (3\pi/2,2\pi) \\
    cos{\theta}-\sin{\theta}\cos{\phi} &= 0, \quad \text{for} \quad \phi \in (\pi/2,3\pi/2)
\end{align}
from which we get the equation of boundary to be,
\begin{align}
    \theta &= -\cot^{-1}{(\cos{\phi})}, \quad \text{for} \quad \phi \in (0,\pi/2) \cup (3\pi/2,2\pi) \\
    \theta &= \cot^{-1}{(\cos{\phi})}, \quad \text{for} \quad \phi \in (\pi/2,3\pi/2)
\end{align}
The key point is that $\theta(\phi)$ is now no longer on a circle. The relevant curve is the intersection of the green surface and the sphere in Fig. \ref{dbl-waist}.
More generally, if we had constructed the union from planes that were at different values of $a$, we could have also gotten $\theta_{asymp}$ values that are not constant everywhere on the curve. In other words, the $\theta_{asymp}$ data on a chosen curve $\theta(\phi)$ is what defines a general spi-subregion.

This observation is general. Here we demonstrated it for a union of two ACDs, but infinite unions can lead to arbitrarily complicated spi-subregions. Note that this structure is similar in AdS as well, once we replace $\theta_{asymp}$ with $\theta_\infty$ as the quantity that has non-trivial structure\footnote{The comparison is for a connected subregion. Note that in AdS4, non-trivial boundary subregions can be described by $\theta_\infty(\phi)$. A non-trivial spi-subregion on Mink$_4$ involves specification of both the curve $\theta(\phi)$ on which $\theta_{asymp}$ is non-vanishing, as well as the values $\theta_{asymp}$. In our simple illustrative example, the curve was non-trivial, but the value of $\theta_{asymp}$ was constant.}. By taking lower dimensional cross sections of  Fig. \ref{fig:acd_union} one can also see that non-trivial unions of spi-subregions result in RT surfaces that include the entire bulk as the cut-off goes to infinity, as was argued in 2+1 dimensions. A further analogy with 2+1 dimensions is that the angle dependence of $\theta_{asymp}$ on a general spi-subregion is in fact realized in 2+1 also, when we note that the two ``ends" of the subregion are the two points\footnote{Note that $S^0$ is a pair of points.} of the $S^0$ that is the analogue of the $S^1$ (the $\phi$-direction) in 3+1.

\section{Ryu-Takayanagi Surfaces in Flat Space}

We argued in the last section that spi-subregions are the natural holographic data that determine RT surfaces in asymptotically flat space. For empty Minkowski space, we know that these RT surfaces are hyperplanes based on general arguments, and we computed their areas. But determining them is a non-trivial task in {\em asymptotically} flat spaces. In this section and the next, we will write the minimal surface equations and solve them perturbatively as well as numerically. We will do this for empty flat space as warm up, and then for black holes. We have discussed the AdS calculations in a form that can be adapted directly to flat space in previous sections, so we will be able to proceed quickly.


\subsection{Empty Flat Space}

RT surfaces in empty flat space are simply lines or (hyper)planes depending on dimension. They have been studied previously from the perspective that the RT data is to be placed on a screen in the bulk \cite{TakayanagiFlat, Qi, Chethan_1}. Our goal in this section and part of Appendix \ref{App:RT} is to formulate the problem in  a more intrinsically holographic way, by using spi-subregion data as the input. We will see eventually that there is a natural connection between the screen radius and the $\Lambda$ of our previous section.

\subsubsection{Mink$_{3}$}

Like AdS$_3$, RT surfaces in $2+1$ dimensional Minkowski space are trivially solvable. The area functional for the co-dimension 2 surface is
\begin{equation}
    \mathcal{A}=\int \sqrt{r^2 + r'(\theta)^2}\; d\theta = \int \underbrace{\sqrt{r^2 \theta '(r)^2+1}}_\mathcal{L} \; dr
\end{equation}
We will treat $\theta$ as a cyclic coordinate as in AdS$_3$. We can write
\begin{equation}
    \frac{d \theta}{d r} = \frac{c}{r\sqrt{r^2-c^2}} \label{eq:EL_eq_4_2}
\end{equation}
where $c$ is an integration constant. Integrating \eqref{eq:EL_eq_4_2} and demanding $r=r_*$ for $\theta=0$ sets $c=r_*$. This is a boundary condition in the bulk. 

Let us make a couple of general comments. Our discussion of RT surfaces for the various cases in this section is done via integrating the equations of motion using data in the bulk (like above) or asymptotic data. This is a standard choice one makes depending on convenience in AdS as well.  In other words an RT surface can determined via angle data at $r \rightarrow \infty$ or closest approach data at $\theta=0$\footnote{Depending on the choice of coordinates one may need to impose a regularity condition in the bulk to get a complete match, as we discussed in the AdS case.}. We mostly solve the system in terms of asymptotic data in the main body of the paper, but present calculations starting from the bulk in Appendix \ref{App:RT}. 

A second point worthy of emphasis is that in both AdS as well as in flat space, our key observation is that the RT surface is determined via $dimensionless$ asymptotic data. In AdS, this is manifest in (say) the Poincare patch where the AdS length scale is simply an overall scale in the metric. The analogous observation in flat space is that the RT surfaces are again determined by angle data as we discussed in the previous section. Note that the structure of the metric is consistent with this fact, where again the scale  $\Lambda$ is  sitting outside expressions like (\ref{eq:conf_met}) and (\ref{conf-factor}).

Coming back to the present discussion, the equation of the curve can be written as
\begin{equation}
    \theta= \cos^{-1}{\left(\frac{r_*}{r}\right)} \label{cos}
\end{equation}
which is just the straight line at a distance $r_*$ from the origin. Substituting it back into the Lagrangian gives us
\begin{equation}
    \mathcal{L}=\frac{r}{\sqrt{r^2-r_*^2}}
\end{equation}
Integrating with proper limits (ie., from $r_*$ to $R_0$) gives us \eqref{eq:area_acd_waist}. These equations, together with \eqref{eq:waist_1} demonstrate that up to the overall conformal rescaling by $\Lambda$ the asymptotic data in the spi-subregion (captured by say $Q$ in our discussion in the previous section) is precisely equivalent to the data in $r_*$. We also learn very explicitly, that the geometry naturally suggests that the scale $\Lambda$ should be chosen to be the screen radius $R_0$.  

\subsubsection{Mink$_{4}$}\label{mink4-curve}

As in AdS, the differential equation is less easily solved in higher dimensions but we can still obtain  power series solutions. This will be enough to establish that the asymptotic data is fixed by the spi-subregion.
For $d=3$ the area functional is,
\begin{equation} \label{eq:area_func_5}
    \mathcal{A}=2\pi \int \underbrace{r\sin{(\theta(r))}\sqrt{r^2 \theta '(r)^2+1}}_\mathcal{L}\; dr
\end{equation}
Note that since a symmetric spi-subregion is symmetric about the North pole, without loss of generality we have assumed the surface also to have the same symmetry. From \eqref{eq:area_func_5} we can write the Euler-Lagrange equation,
\begin{equation}\label{eq:EL_eq_5}
        \frac{\partial }{\partial \theta}\left(r\sin{(\theta(r))}\sqrt{r^2 \theta '(r)^2+1}\right)=\frac{d}{d r}\frac{\partial }{\partial \theta'}\left(r\sin{(\theta(r))}\sqrt{r^2 \theta '(r)^2+1}\right)
\end{equation}
Like in AdS we can try a power series expansion of for large values of $r$ in the form\footnote{Again as in AdS, to keep the dimensions straight, we need to work with an expansion in $\frac{r}{\Lambda}$ rather than $r$. It is a simple but important fact that it is precisely this data that is fixed by the spi-subregion, cf. \eqref{eq:waist_1}.}
\begin{equation}\label{eq:exp_2}
        \theta(r)=\sum_{i=0}\frac{\theta_i}{r^i}
\end{equation}
Substituting \eqref{eq:exp_2} in \eqref{eq:EL_eq_5} and comparing coefficients we get the solution to be,
\begin{equation}
    \theta(r)=\frac{\pi }{2}+\frac{\theta_1}{r}+\frac{\theta_1^3}{6 r^3}+\frac{3 \theta_1^5}{40 r^5}+\frac{5 \theta_1^7}{112 r^7}+\dots
\end{equation}
As expected, this is just the expansion of \eqref{cos}  which is the equation of a plane in 3+1 dimensions. It follows therefore from \eqref{eq:waist_1} that the data that determines the RT surface, $\theta_1$, is nothing but the spi-subregion data, $\tan Q$. Substituting the solution back in the Lagrangian we get,
\begin{equation}\label{eq:lag_3}
    \mathcal{L}= r
\end{equation}
Integrating with proper limits gives us again back \eqref{eq:area_acd_waist} for $d=3$.

\subsubsection{Mink$_5$}

Again the differential equation is not directly solvable but we can still obtain a power series solutions from the area functional
\begin{equation}
    \mathcal{A}=\int \underbrace{r^2\sin^2{(\theta(r))}\sqrt{r^2 \theta '(r)^2+1}}_\mathcal{L} \;dr
\end{equation}
and the corresponding equations:
\begin{equation}
        \frac{\partial }{\partial \theta}\left(r^2\sin^2{(\theta(r))}\sqrt{r^2 \theta '(r)^2+1}\right)=\frac{d}{d r}\frac{\partial }{\partial \theta'}\left(r^2\sin^2{(\theta(r))}\sqrt{r^2 \theta '(r)^2+1}\right)
\end{equation}
A new feature is that substituting the power series expansion and solving for the coefficients gives us an extra integration constant which we call $\theta_2$:
\begin{equation}
    \theta(r)=\frac{\pi }{2}+\frac{\theta_1}{r}+\frac{\theta_2}{r^2}+\frac{\theta_1^3}{6 r^3}+\frac{\theta_1^2 \theta_2}{r^4}+\dots \label{eq:mink_5sol}
\end{equation}
Substituting it in the Lagrangian we get,
\begin{equation}\label{eq:lag_3}
    \mathcal{L}=r^2-\frac{\theta_1^2}{2}+\frac{\theta_2^2-\frac{\theta_1^4}{8}}{r^2}+\frac{3 \theta_1^2 \theta_2^2-\frac{\theta_1^6}{16}}{r^4}+\dots
\end{equation}

The extra constant $\theta_2$  is a feature\footnote{The extra freedom  in $\theta(r)$ appears at  $\mathcal{O}(r^{-(d-2)})$ in $d+1$ dimensions. In $d=3$ this requires a $\log$ term in the series expansion to get the most general solution. This is explained more completely in Appendix \ref{cylinder}.} of all $d >3$. Setting it to zero leads to a match with our expectation that the curve is a hyperplane (ie., \eqref{cos}) and that the area integral functional matches with \eqref{eq:area_acd_waist}. As in AdS the extra integration constant is to be understood as a regularity condition one needs to impose in the bulk. We solve the same system in the $r(\theta)$ language, to illustrate this in Appendix \ref{App:RT}.

\subsection{Black Hole in the Bulk}

A key advantage of the spi-subregion approach is that it allows us to cleanly separate out the asymptotic dimensionless information that controls the RT surface, from the scale information. In particular, we can now repeat the RT surface calculation above, in any asymptotically flat spacetime using the same procedure. This approach is to be contrasted with trying to define data on a finite radius screen. In the latter approach, one has to worry about  (eg.,) the induced metric on the screen -- the canonical data becomes less clear. Perhaps because of this, even though there have been papers on RT surfaces in empty flat space anchored to screens (ie., the hyperplanes we saw earlier\footnote{Note however that we worked them out in terms of data at spi and not in terms of screen data directly.}) much less effort has been directed towards RT surfaces of flat space black holes. In this section and Appendix \ref{App:RT}, we will construct these black hole RT surfaces perturbatively (both around the asymptotic region and around the deepest point in the bulk) and numerically (in the entire spacetime). Since there are no black holes in 2+1 dimensional flat space, we start with 3+1 dimensions.

\subsubsection{Schwarzchild$_4$}

We start from the Schwarzchild metric in 3+1 dimensions
\begin{equation}
    ds^2=-\left(1-{\frac{r_{\mathrm{s}}}{r}}\right)dt^{2}+\left(1-{\frac {r_{\mathrm {s}}}{r}}\right)^{-1}dr^{2}+r^{2}d\Omega_2 ^{2} \label{Sch4}
\end{equation}
\begin{figure}[h!]
    \centering
    \includegraphics[height=10cm]{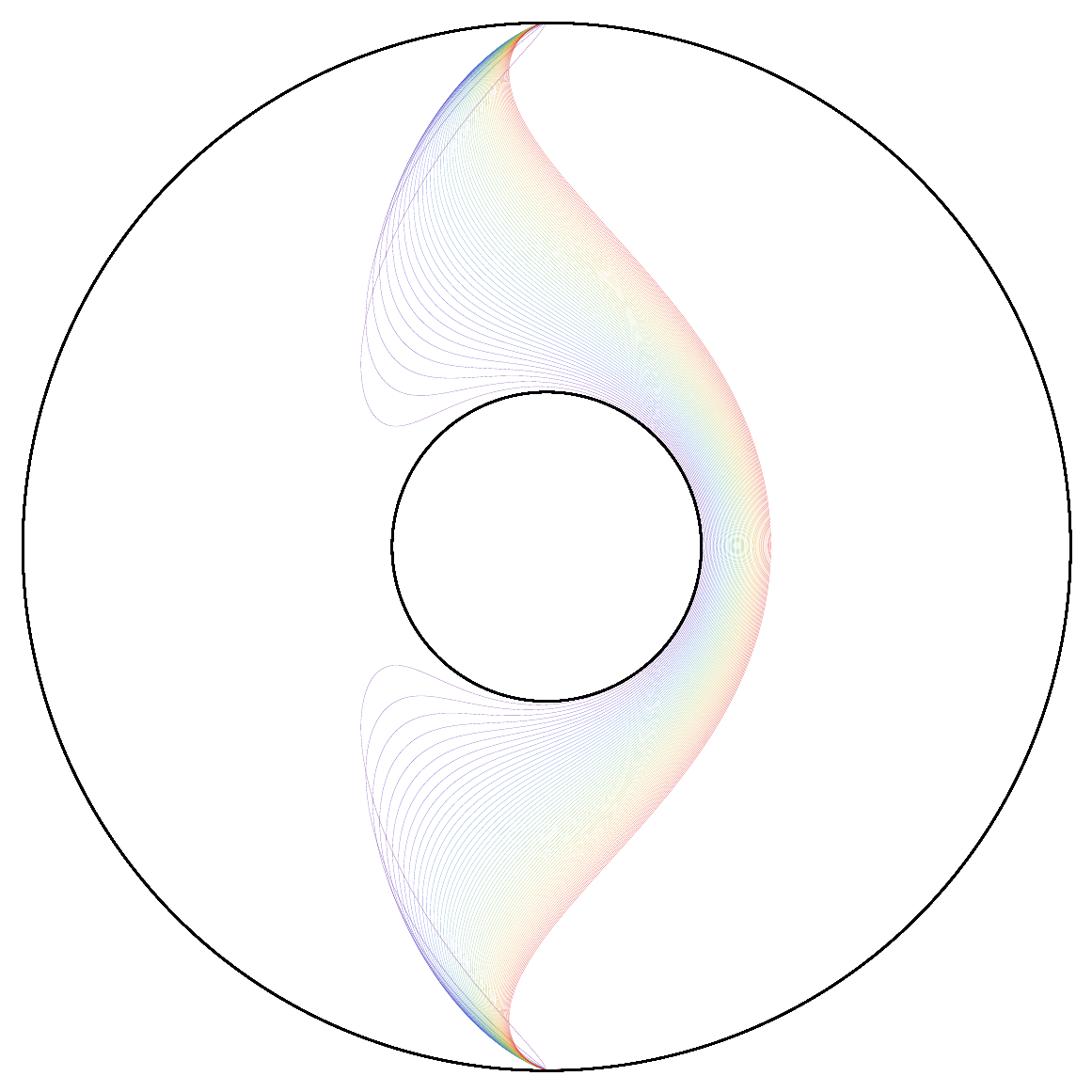}
    \caption{Plots of RT surfaces on Schwarzschild$_4$ in (asymptotically) conformal coordinates as the closest point of approach is varied. The plots are for $r_s = 0.5$. The curve of farthest approach (the reddest curve) is chosen arbitrarily, there is no interesting information to be gathered by going further to the right. The bluest curve will be discussed further in the next section.}
    \label{fig:BH_3+1_RT_surface}
\end{figure}
Without loss of generality we can assume the RT surface associated to the symmetric spi-subregion to be symmetric about the North pole. Then the area functional of a co-dimension 2 surface is,
\begin{equation}
    \mathcal{A}=2\pi\int \underbrace{r \sin{\theta} \sqrt{\left(1-{\frac {r_{\mathrm {s}}}{r}}\right)^{-1} +r^{2}\theta'^2}}_\mathcal{L} \;\mathrm{d}r\label{eq:3.2}
\end{equation}
The Euler-Lagrange equations yield
\begin{equation}
    \frac{\mathrm{\partial}}{\mathrm{\partial}\theta}\left( r \sin{\theta} \sqrt{\left(1-{\frac {r_{\mathrm {s}}}{r}}\right)^{-1} +\theta'^{2}r^{2}}\right)=\frac{\mathrm{d}}{\mathrm{d}r}\frac{\mathrm{\partial}}{\mathrm{\partial}\theta'}\left(r \sin{\theta} \sqrt{\left(1-{\frac {r_{\mathrm {s}}}{r}}\right)^{-1}+ \theta'^{2}r^{2}}\right)\label{eq:el_eq}
\end{equation}
As before, assuming a power series solution of the form
\begin{equation}\label{eq:exp_4}
    \theta(r)=\sum_{i=0}\frac{\theta_i}{r^i}
\end{equation}
leads to
\begin{equation}
    \theta(r)=\frac{\pi }{2}+\frac{\theta_1}{r}-\frac{r_s \theta_1}{2 r^2}+\frac{\frac{2 \theta_1^3}{3}-\frac{r_s^2 \theta_1}{2}}{4 r^3}+\frac{-3 r_s^3 \theta_1-28 r_s \theta_1^3}{48 r^4}+\dots
\end{equation}
Substituting this into the Lagrangian of \eqref{eq:3.2} we get,
\begin{equation}
        \mathcal{L}=r+\frac{r_s}{2}+\frac{3 r_s^2}{8 r}+\frac{5 r_s^3-16 r_s \theta _1^2}{16 r^2}+\frac{5 \left(7 r_s^4+16 r_s^2 \theta _1^2\right)}{128 r^3}+ \dots
\end{equation}

\begin{figure}[h!]
    \centering
    \includegraphics[width=17cm]{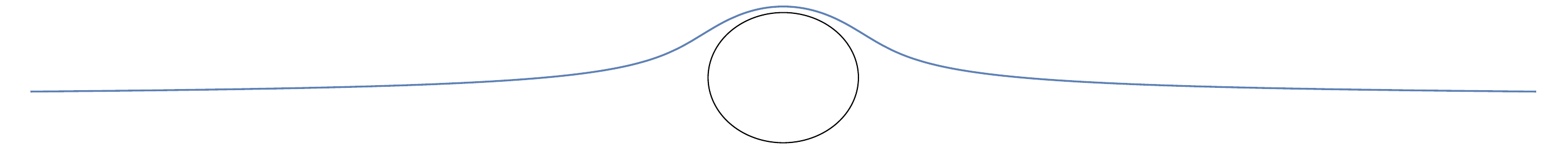}
    \caption{A sample plot of the RT surface in physical coordinates.}
    \label{physical-coord}
\end{figure}

An analogous perturbative solution around the point of nearest approach to the horizon is presented in Appendix \ref{App:RT}. One can in fact, do more. The ODE in \eqref{eq:el_eq} can be numerically solved with the boundary conditions
\begin{eqnarray}\label{eq:bound_cond}
        r(\theta)\Big|_{\theta=0} = \;r_* ,  \  \ \  
        r'(\theta)\Big|_{\theta=0} =\;0    
\end{eqnarray}
Plotting the numerical solutions, we arrive at Fig \ref{fig:BH_3+1_RT_surface}. The figures have been plotted in compact coordinates $(\rho, \phi)$ with the following identification
 \begin{equation}
     \rho = \tanh{(r)}, \qquad \phi = \theta.
 \end{equation}
In Fig \ref{fig:BH_3+1_RT_surface} the RT surfaces start on the right side of the black hole, wrap around to the other side and then turn around to reach the antipodal points. Of course in non-conformal coordinates, this simply corresponds to the curves becoming asymptotically planar, see Fig. \ref{physical-coord}.
At the turning point the slope blows up. So, we also need to solve the equations by interchanging the independent and dependent variable in the action functional in order to fully determine the curve.

\subsubsection{Schwarzchild$_5$}

A similar calculation can be done in higher dimensions also. The results are also largely parallel -- except for the fact that there is the extra integration constant, when we perturbatively solve the system at large $r$. This is the same integration constant that we found in empty 4+1 dimensional Minkowski space, and again the explanation is that it is an artifact of the absence of manifest regularity in the bulk. 

The metric for 4+1 Schwarzchild black hole is,
\begin{equation}
    ds^2=-\left(1-{\frac{r_{\mathrm{s}}^2}{r^2}}\right)dt^{2}+\left(1-{\frac {r_{\mathrm {s}}^2}{r^2}}\right)^{-1}dr^{2}+r^{2} d\Omega_3^2 \label{Sch5}
\end{equation}
\begin{figure}[h!]
    \centering
    \includegraphics[height=10cm]{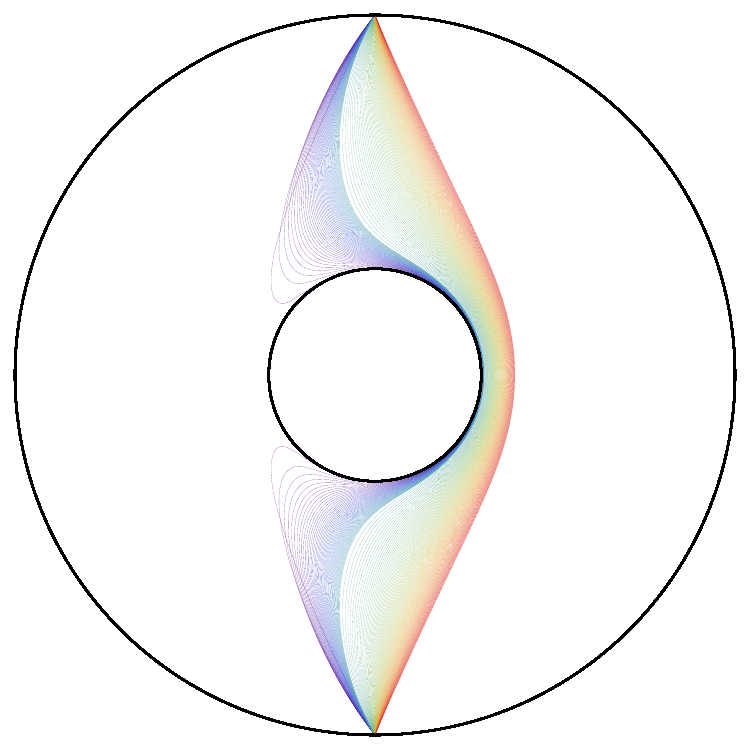}
    \caption{Plot of minimal surfaces in Schwarzschild$_5$ in conformal coordinates. Similar comments as in Fig. \ref{fig:BH_3+1_RT_surface} apply.}
    \label{fig:conf_4}
\end{figure}
Again without loss of generality we are assuming the surface to be symmetric about the pole. The area functional of the co-dimension 2 surface is
\begin{equation}\label{eq:area_func_flat_5}
        \mathcal{A}=4\pi \int \underbrace{r^2 \sin^2{\theta(r)} \sqrt{\left(1-{\left(\frac {r_{\mathrm {s}}}{r}\right)^2}\right)^{-1} +\theta'(r)^2r^{2}}}_\mathcal{L}\; dr
\end{equation} 
and the equations of motion are 
\begin{equation}
    \frac{\mathrm{\partial}}{\mathrm{\partial}\theta}\left(r^2 \sin^2{\theta(r)} \sqrt{\left(1-{\left(\frac {r_{\mathrm {s}}}{r}\right)^2}\right)^{-1} +\theta'(r)^2r^{2}}\right)=\frac{\mathrm{d}}{\mathrm{d}r}\frac{\mathrm{\partial}}{\mathrm{\partial}\theta'}\left(r^2 \sin^2{\theta(r)} \sqrt{\left(1-{\left(\frac {r_{\mathrm {s}}}{r}\right)^2}\right)^{-1} +\theta'(r)^2r^{2}}\right) \label{ODE}
\end{equation}
The power series solution takes the form
\begin{equation}\label{eq:bh_rt}
    \theta (r)=\; \frac{\pi }{2}+\frac{\theta_1}{r}+\frac{\theta_2}{r^2}+\frac{\theta_1^3-3 r_s^2 \theta_1}{6 r^3}+\dots
\end{equation}
On substituting this $\theta(r)$ back into the Lagrangian of \eqref{eq:area_func_flat_5} we get,
\begin{align}\label{eq:lag_4}
    \mathcal{L}= r^2 + \frac{1}{2} \left(r_s^2-\theta_1^2\right)+\frac{3 r_s^4-10 r_s^2 \theta_1^2-\theta_1^4+8 \theta_2^2}{8 r^2} + \dots
\end{align}

A similar perturbative calculation where the dependent and independent variables have been reversed, is presented in \ref{App:RT}. Just as in 3+1 dimensions, the ODE in \eqref{ODE} can be numerically plotted with similar boundary condition as in \eqref{eq:bound_cond}. Plotting the solution we get Fig \ref{fig:conf_4}. 

Some comments are in order. Firstly, if we try to calculate the distance of the RT surface \eqref{eq:bh_rt} from the $r \cos{\theta}=0$ surface, we get
\begin{equation}
    r\cos{(\theta(r))} \approx -\theta_1+\frac{r_s \theta_1}{2 r}+\frac{r_s^2 \theta_1}{8 r^2}+ \frac{\frac{r_s^3 \theta_1}{16}+\frac{r_s \theta_1^3}{3}}{r^3}+\frac{\frac{5 r_s^4 \theta_1}{128}-\frac{43 r_s^2 \theta_1^3}{96}}{r^4}+ ...
\end{equation}
from which we can see that as $r\rightarrow\infty$ the distance becomes constant i.e. $r\cos{\theta}\sim-\theta_1$. This implies that RT surface becomes a hyperplane at large distance from the horizon. This distance acts as the boundary data as we saw in the empty Minkowski case. Comparing to \eqref{eq:waist_1} we get
\begin{equation}
    \theta_1= \tan{Q}.
\end{equation}
Note that in this result and indeed in the perturbative expansions above, we are again implicitly working with $\frac{r}{\Lambda}$ as the variable and not $r$. This is what ensures that the $\theta_i$ are dimensionless variables. It also means that once $\frac{r_s}{ \Lambda}$ replaces $r_s$ in the various expressions, the philosophy of the calculations is identical to that in the empty Minkowski case. Of course, this is simply an illustration that the asymptotic data, even with black holes, is captured by spi-subregions.

Let us repeat an important observation. In the AdS case, the divergent terms of the Lagrangian in the area functional were independent of the black hole radius. In fact, a stronger statement was true -- the divergent terms in the AdS black hole case are the same as those in empty AdS\footnote{Because of this, one sometimes suppresses the divergences in the area by subtracting \eqref{eq:lag_1} from \eqref{eq:lag_2}, when one is interested in {\em comparing} the entanglement entropies.}. In essence, what this is showing is that the divergence structure of the entanglement entropy is {\em independent} of the state in AdS/CFT. This is natural because the holographic dual of AdS gravity is a quantum field theory, and the entanglement entropy of a quantum field theory subregion has a universal short distance divergence.  

But we see that such a prescription will not work in flat space -- \eqref{eq:lag_4} has an extra divergent term compared to \eqref{eq:lag_3} which depends on the black hole radius. This is consistent with the suggestion that the holographic dual of flat space quantum gravity is {\em not} a local quantum field theory. We suspect that the precise expression of the result above, may be useful in deciphering the form of flat space hologram.



\section{Ryu-Takayanagi Phase Transitions}

\begin{figure}[h]
    \centering
    \includegraphics[width=10cm]{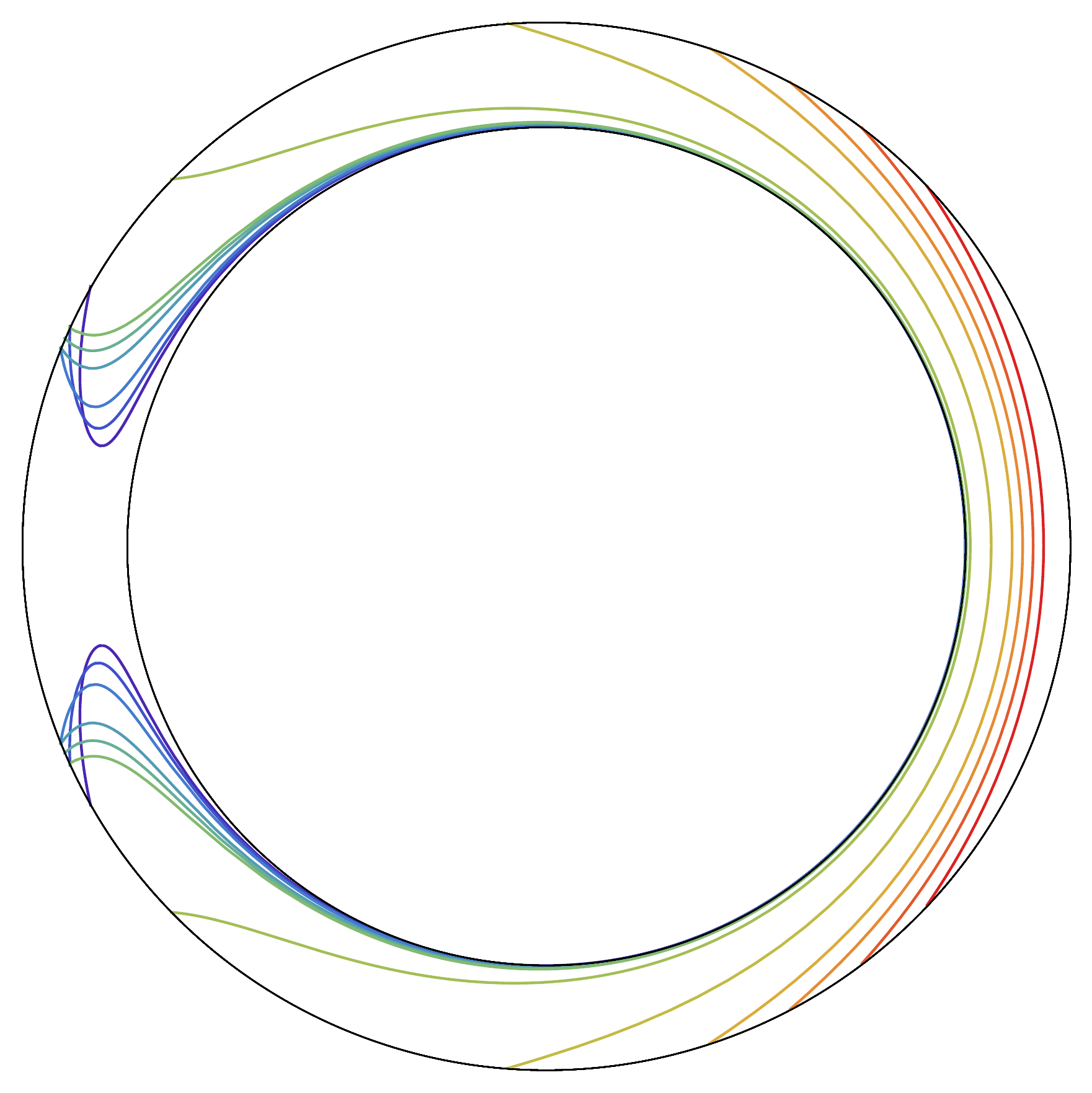}
    \caption{A plot of RT surfaces with horizon and a cut-off screen. In this plot for clarity we have taken $r_s=80$ and $R_0=r_{\text{cut}}=100$. As the curves turn from red to blue, they get closer to the horizon -- this means initially that the spi-subregion size is increasing, but eventually it starts decreasing. This is clear from this figure, as well as the extreme blue curve in  Fig \ref{fig:BH_3+1_RT_surface}.}
    \label{fig:BH_cut}
\end{figure}

So far we have looked in detail at the geometry of RT-surfaces in asymptotically flat spaces, given the boundary data $Q$. In this section we will present how the RT-surface behaves as we change the ``size'' of this spi-subregion when there is a black hole in the bulk. For a more intuitive understanding of this section it may be useful to take a look at the conformal plot Fig \ref{fig:BH_3+1_RT_surface}, a version of which we repeat in Fig \ref{fig:BH_cut} (with emphasis on the cut-off and some other salient features).

We start from a small spi-subregion -- these correspond to the redder curves in Fig \ref{fig:BH_3+1_RT_surface}. These curves do not penetrate too deep into the bulk and are very similar to the empty Minkowski curves as the space is almost flat far from black hole. Now, as the spi-subregion size increases the curves penetrate deeper into the bulk and effects of curvature become more prominent. One can see in Fig \ref{fig:BH_3+1_RT_surface} that as the asymptotic angle of approach (and therefore the spi-subregion size) increases the RT surfaces move closer to the black hole and start to wrap around the horizon. But there is an interesting point to note here - the most extreme blue curve is closest to the black hole, but has a smaller asymptotic angle than the previous curve.  In other words, as the RT surfaces go deeper into the bulk   beyond a certain point, the asymptotic angle of approach (and therefore the spi-subregion size) starts to decrease again.
\begin{figure}[h!]
\centering
     \begin{subfigure}[t]{0.4\textwidth}
            \centering
            \includegraphics[width=8cm]{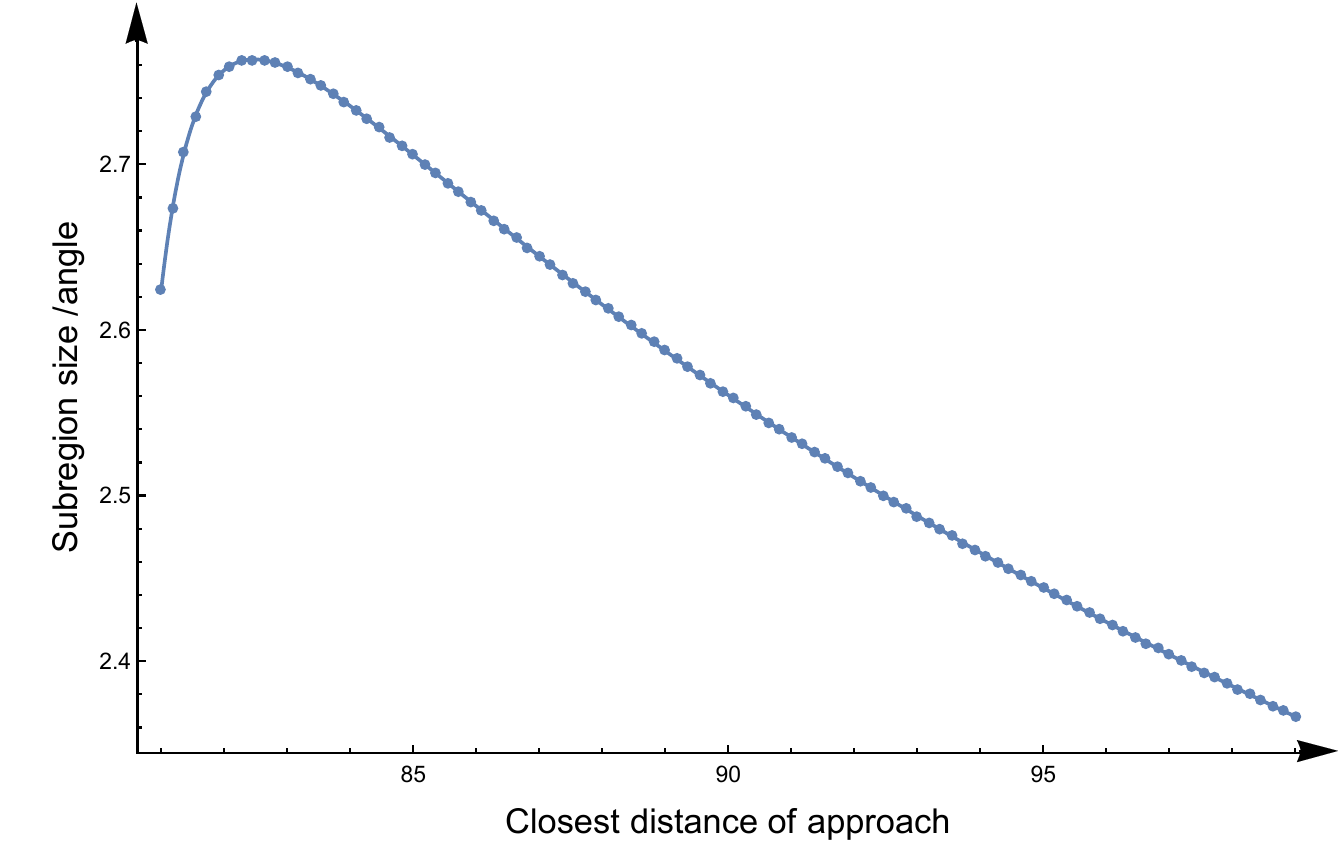}
            \caption{}
            \label{fig:RT_plot}
     \end{subfigure}
     \hfill
      \begin{subfigure}[t]{0.55\textwidth}
            \centering
            \includegraphics[width=8cm]{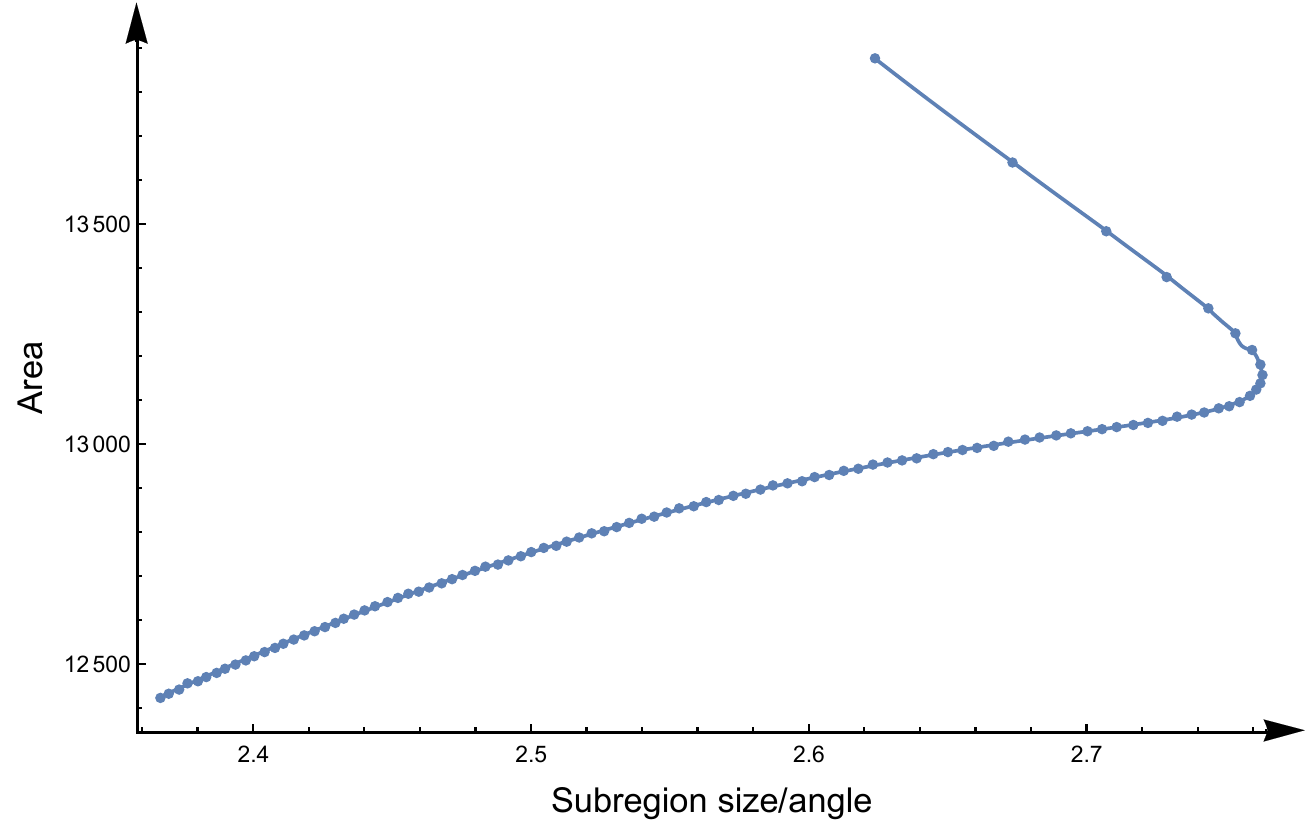}
            \caption{}
            \label{fig:area_size}
     \end{subfigure}  
     \caption{(a) Plot of the subregion size (taken to be the angle subtended by the intersection of the RT-surface and cut-off screen) vs the closest distance between the RT-surface and the black hole. The vertical axis is in radians and the horizontal axis is measured from origin, with $r_s=80$. (b) Plot of the area of the RT-surface (computed as the integral of \eqref{eq:3.2} from distance of closest approach to $r_{cut}=100$) vs subregion size, zoomed in near the transition.  We see that for the same subregion we have two surfaces -- one with higher area is closer to the black hole and therefore we discard these surfaces. }
     \label{fig:plots}
\end{figure}

To better understand the geometry and physics, it is convenient to put a cut-off screen at a finite distance from the black hole. Then the subregion size induced on the screen can be used as a proxy for the size of the spi-subregion\footnote{Note that this is merely for convenience in visualization. The holographic data is of course defined at spi.}. Doing this we get the plot in Fig \ref{fig:BH_cut}. We can see that as the surface moves closer to the black hole, the boundary subregion size increases but only upto a critical distance -- beyond which if we move any closer, the subregion size begins to decrease. 

The size of the subregion has been plotted against the distance of closest of approach to the horizon in Fig \ref{fig:RT_plot}. It is clear from the plot that for a given subregion we may have more than one extremal surface. The RT prescription suggests that we resolve this by selecting the surface with the smallest area. The plot of area vs subregion size is presented in Fig \ref{fig:area_size}. The higher area after the turnaround comes from the RT-surfaces closer than a critical distance. Therefore the surfaces which are farther from the horizon than the critical distance are the acceptable RT surfaces. This structure is identical to that in AdS.

\begin{figure}[h!]
    \centering
    \includegraphics[width=11cm]{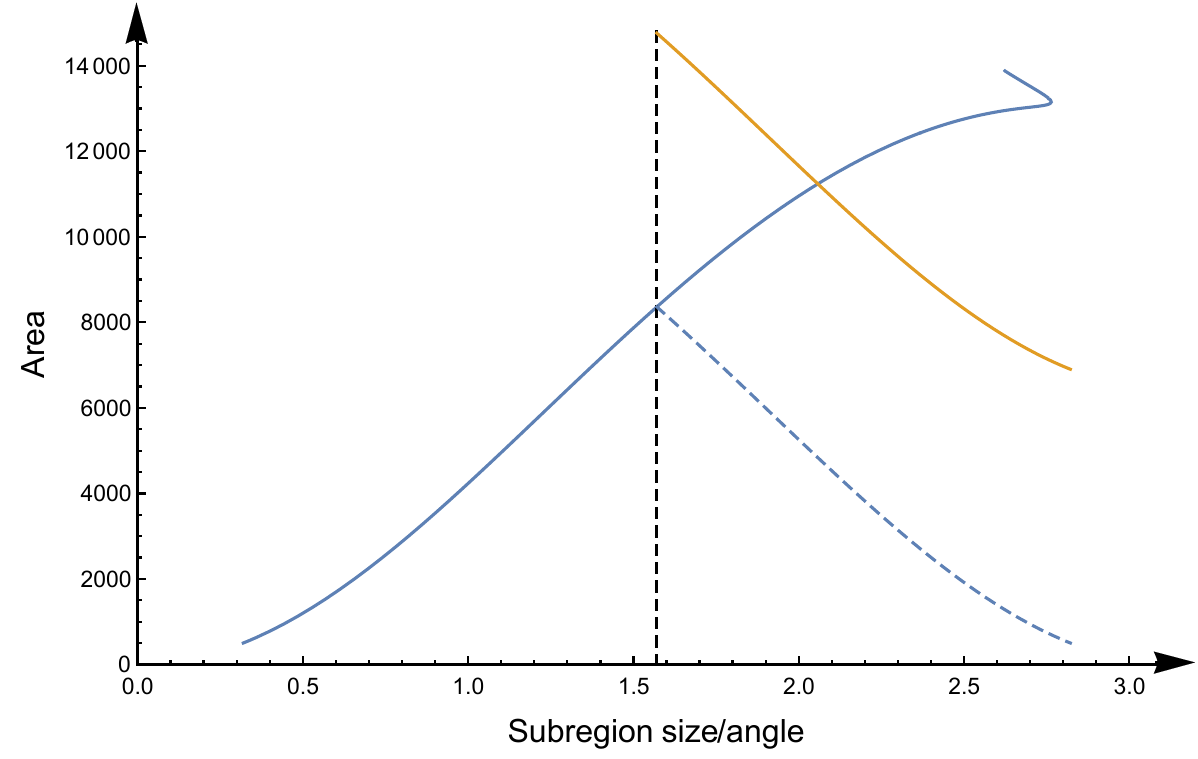}
    \caption{Plot of area of RT surface vs subregion size, but for the entire range of subregion sizes. The black dashed line corresponds to the $\pi/2$ subregion on the screen. The blue dashed line corresponds to the RT surface for the complementary subregion. The orange line is the area corresponding the disconnected RT-surface. It is shifted by a constant (equal to the horizon area) from the dashed blue curve. It is at the point of intersection of the blue and the orange lines the we transition from a connected RT-surface to a disconnected RT-surface that contains the horizon. }
    \label{fig:area_size(2)}
\end{figure}

As in AdS, there is one more catch -- disconnected surfaces are also acceptable RT surfaces as long as they are homologous to the subregion. So, one needs to check for such RT-surfaces which have smaller areas than the connected surfaces. We have plotted the areas of the relevant surfaces in Fig \ref{fig:area_size(2)}. The point of intersection between the blue line and the orange line shows the transition between connected and disconnected surfaces. Beyond this point the RT-surface consists of the black hole horizon together with the minimal surface corresponding to the complement of the boundary subregion under consideration. This also suggests that it is natural to view the RT-surface for the full boundary (or spi) as the black hole horizon. Recall that this is also the case with AdS black hole geometries. 

The picture we have here is very similar to that in AdS, once we make an analogy between  $r_{cut}$ and the AdS length scale\footnote{This is another hint that the screen radius is best viewed as an IR cutoff in flat space.}. Usually in AdS, the discussion of black hole RT surface phase transitions is made in the context of large black holes, but the discussion can be extended to small black holes as well \cite{Freivogel}. There are two qualitative distinctions between small and large AdS black holes and their RT surfaces. The first is that for small black holes the phase transition between the connected and disconnected RT surfaces happens closer  and closer to $\theta_\infty = \pi/2$ as the black hole becomes smaller and smaller (ie., $r_s/L \rightarrow 0$). The second is that for small black holes the distance of closest approach of the transition RT surface to the horizon is comparable to the size of the horizon\footnote{For large AdS black holes the closest approach can be much smaller \cite{Freivogel}.}. We have checked that both these statements are true for flat space black holes as well, as long as we replace $L$ with $r_{cut}$. In other words the RT phase transition structure of both small and large AdS black holes have natural analogues in flat space as well. 


Let us make some comments about the screen \cite{BudhadityaGeneral}. Note that in AdS, $L$ provides the IR scale while the  radial cut-off provides the (UV) regulator for the divergence in the entanglement entropy/RT-surface area. Here on the other hand, the screen radius doubles both as the IR scale and regulator, and there are no UV divergences. These observations are a strong indication that the choice of an IR scale and its interpretation is likely of fundamental significance in flat space quantum gravity. It was suggested in \cite{CriticalIslands, DirichletBaths} that the screen size may be related to a coarse-graining scale. 

\section{Comments on the Dual Interpretation}

Our results above paint a rich and suggestive  picture on the bulk side of the holographic duality. We learnt that RT surfaces in flat space have a geometric understanding in terms of data at spatial infinity and that the bulk cut-off has an IR interpretation which is natural from multiple perspectives. If holography in flat space is more than just fiction, it is reasonable to suspect that these structures compute some form of entanglement entropy\footnote{Even though not widely explored, we find it possible that flat space quantum gravity should perhaps be treated as an open system. In such a scenario, the mixed density matrices of interest need not always be reduced density matrices within a unitary theory, and we will be interested in more general von Neumann entropies as well. But we will refer to these also loosely as entanglement entropies here, by exploiting the possibility that they can be purified by considering also the bath.} in the microscopic theory. But since an intrinsic definition of the flat space hologram is unknown\footnote{We will briefly mention two distinct (but possibly related) directions that have been proposed in the quest for flat space holography, in the next section.}, we will not be able to make a direct statement about {\em how} the entanglement (entropy) is to be computed\footnote{But see the discussion in the last  paragraph of this section.}. In this section we will instead give strong evidence that the regulated areas of the RT surfaces we have identified, compute {\em some} entanglement entropy. Precisely because of the unavailability of a sufficiently explicit hologram, this needs a justification.

Let us step back for a moment and re-consider the AdS case: why would we think that we are dealing with entanglement (entropy), if we were only aware of {\em bulk} calculations of RT surfaces in AdS\footnote{Note that even in AdS, apart from the case of AdS$_3$/CFT$_2$, direct CFT comparisons of entanglement entropy are unavailable --  weakly coupled gravity (where areas of surfaces make sense) is dual to strongly coupled gauge theory.}? To get some context, we start by listing some features of AdS RT surfaces.
\begin{itemize} 
\item RT surfaces are defined by data on the conformal boundary, and this data allows the interpretation of a bi-partitioning. 
\item The areas of AdS RT surfaces defined this way  satisfy a whole list of entanglement entropy inequalities (see eg., \cite{VyshnavVestige}, an original reference is \cite{Headrick}). These inequalities are highly non-trivial to prove in conventional quantum mechanics (eg., proof of strong subadditivity \cite{Lieb}), but are simple and intuitive geometric statements from the AdS bulk perspective.
\item Since the hologram of AdS is a local theory, we would expect it to exhibit area scaling of entanglement entropy. The divergent areas of the AdS RT surfaces indeed scale with a codimension-2 divergence (we are counting boundary dimensions here). 
\item The conventional subregion-subregion duality in AdS is based on constructing quantum extremal surfaces (or RT surfaces in the classical limit) to define entanglement/causal wedges. This idea has recently had a remarkable renovation in terms of subalgebra-subregion duality, where bulk subregions which are causally complete have been argued to be dual to emergent type III$_1$ algebras in the boundary theory in the large-$N$ limit \cite{LeutheusserNew}. These ideas provide a natural generalization of the notion of a boundary subregion. A key message of these developments is that causally complete bulk regions are associated to large-$N$ divergences in entanglement in the holographic dual. 
\item It has been argued (and proved in various toy models) \cite{Harlow} that the existence of an RT formula for entanglement entropy is equivalent to the quantum error correction (QEC) property of bulk reconstruction \cite{ADH}. This suggests that the ``additivity anomaly" \cite{LeutheusserNew} and related union/intersection properties of bulk regions within RT surfaces, are indicators of entanglement.
\end{itemize}
There are of course many other related ideas, our goal above is simply to sample some key features. We emphasize that in many of the above statements, the finiteness of the entropy/area and therefore the introduction of a bulk cut-off, is crucial. 

In the following, we point out that {\em all} of these bullet points have natural analogues in flat space:
\begin{itemize}
\item A key result of the present paper is that the first bullet point above, is true even in asymptotically flat space. This is a technical result -- our motivation was to clarify that the data describing an RT surface in flat space can be captured by {\em asymptotic} data at spi. Flat space RT surfaces in terms of data on screens have been explored before, see eg. \cite{TakayanagiFlat, Qi}. But this is not fully satisfactory because it sacrifices some of the spirit of holography. The results of this paper clarify that asymptotic data can do the job in flat space as well. This makes the bulk description of RT surfaces in flat space, as legitimate (ie., as holographic) as in AdS with the screen radius acting as a regulator in both cases.
\item As we noted above, there is a whole slew of highly non-trivial entanglement entropy inequalities that are valid in AdS. Remarkably, there is strong evidence that {\em all} of these are  also valid in flat space in terms of regulated sizes of minimal surfaces \cite{VyshnavVestige}. This seems to not be as widely appreciated as it should be, even though the proofs are entirely trivial in 2+1 dimensions \cite{VyshnavVestige}\footnote{In fact it was suggested in \cite{VyshnavVestige} that these may be the simplest proofs of these entropy inequalities in existence. The original quantum mechanics proof of eg. strong sub-additivity \cite{Lieb} is quite non-trivial. The AdS proofs in eg., \cite{Headrick} are significantly simpler, but the proofs in \cite{VyshnavVestige} can legitimately be called trivial, because they follow from (sometimes elaborate) applications of triangle inequality.}. In higher dimensions, the proofs are technically more difficult because of the complicated bulk structures that arise from unions and intersections. (This is true even in AdS, and the complications in flat space are structurally analogous and should be simpler.) These observations are a strong suggestion that the objects computed by these flat space RT surfaces should indeed be interpreted as entanglement entropy.
\item The hologram of flat space is believed to be a non-local theory and it is a typical expectation that such theories exhibit volume scaling of entanglement entropy and not area scaling. The intuition is that the vacua (and excited states) of  highly non-local theories are much closer to typical states in the Hilbert space, than those of local theories\footnote{This expectation has been borne out in theories like the SYK model, where the entanglement entropies of sub-systems exhibit volume scaling \cite{SYK-vol}.}. Typical states under bi-partition lead to volume scaling of the entanglement entropy, a fact that follows from Page's theorem \cite{Page}. Such volume scaling is indeed what we find in our flat space RT computations as well, where the leading divergence is codimension-1 as opposed to codimension-2. 
\item  According to subalgebra-subregion duality in AdS/CFT \cite{LeutheusserNew}, boundary sub-algebras\footnote{These type III$_1$ sub-algebras subsume the notion of a boundary subregion -- the older language of subregions is what we have mostly followed in this paper, except in the present discussion.} in the large-$N$ (ie., semi-classical) limit are identified with causally complete subregions of the bulk. In flat space, we certainly have causally complete {\em bulk} subregions\footnote{Causal developments of bulk subregions, (asymptotic) causal diamonds, etc. are examples of causally complete bulk subregions.}, but the dual theory is more mysterious. 
 The results of \cite{LeutheusserNew} strongly suggest that one should associate large-$N$ type III$_1$ subalgebras as the natural structure underlying spi-subregions as well. Crucially, in both AdS and flat space, the $N \rightarrow \infty$ limit can be taken while keeping the bulk cut-off fixed\footnote{It is important here that the type III structure is attributed to the large-$N$ limit and not the continuum limit \cite{LeutheusserNew}.}, and we expect sharp bulk causal structures to emerge in this limit.
 
A noteworthy feature\footnote{Some of these ideas are currently being developed and we hope to report on them elsewhere.} of the algebra language is that the causal structure of the {\em boundary} theory is not as directly significant in determining the bulk causal structure, as one might have naively thought. (It is tempting to think that the absence of a causal structure at the conformal boundary of flat space is another manifestation of this large-$N$ feature.) This is because time evolution on the boundary is sensitive to the full operator algebra of the gauge theory, but the large-$N$ limit suppresses everything except the single trace sector. This was crucial for the discussions in \cite{LeutheusserNew} who show that bulk casual diamonds that are disconnected from the boundary can be encoded in the algebras of time bands on the boundary. Note that time bands around the $t=0$ slice of spatial infinity are automatically present in the boundary of flat space as well, and they are related to bulk causal diamonds in a manner identical to that in AdS. With spi-subregions as duals of ACDs and time bands as duals of bulk causal diamonds, we are lead to a picture that is naturally isomorphic to Fig. 1 of \cite{LeutheusserNew}. The claim of the duality would be that there are natural type III$_1$ algebras that can be associated to these boundary regions. Clearly this is an important hint about the entanglement structure of the non-local theory that describes flat space gravity in the large-$N$ limit. 

A related comment is that as long as we are in the large-$N$ limit, it may be possible to get insights about black holes in flat space, by exploiting their connections to small black holes in AdS. The possibility that small black holes in AdS are dual to excited $M \times M$ submatrix configurations of a large-$N$ theory with $ M < N$ (with $M/N$ held fixed in the large-$N$ limit\footnote{Note that in the strict large-$N$ limit, small black holes of this type do not evaporate.}) has been proposed before \cite{Asplund, Hanada, Chethan_2}. It remains to be seen if the algebra language is instructive in studying such configurations. The case studied in \cite{LeutheusserFirst} was for large black holes above the Hawking-Page transition, but it seems to us that this may be inessential as long as the state has (a) a black hole interpretation, and (b)  is classically stable (even if not thermodynamically). Note in particular that one of the crucial arguments for emergent type III$_1$ in \cite{LeutheusserFirst} came from the spectrum of the Hartle-Hawking correlator, which in turn is obtained by solving the bulk wave equation with infalling boundary conditions. Even though the explicit form of the small black hole metric is only known numerically \cite{Santos}, the spectrum statement will hold true since the natural boundary conditions are infalling. 

If we step away from the large-$N$ limit, we expect these black holes to evaporate. We expect the bulk discussion of the flat space information paradox to proceed without much difference from the familiar AdS-coupled-to-bath system -- with the AdS length scale replaced by the IR cut-off here\footnote{Just as in AdS we also have to replace RT surfaces with quantum extremal surfaces (QES). Note that the bulk definition of QES is identical in flat space and AdS, the difference lies merely in that we are using (spi-)subregions.}. Note that the various technicalities we had to deal with in the earlier sections of the paper are all related to (spi-)subregions, and these do not show up in the discussion of evaporation. This is because we are interested in the entanglement wedge of the full boundary theory and not that of a proper spi-subregion when discussing the information paradox.

\item Finally, let us mention that the union-intersection structure associated to quantum error correction was already noted for flat space in \cite{Chethan_1}. In the context of subalgbra-subregion duality, this is simply a version of the ``additivity anomaly" \cite{LeutheusserNew}. Our observations about non-trivial unions of distinct ACDs is nothing but an observation about the nature of the additivity anomaly in flat space. 
\end{itemize}

Before we close this section, let us emphasize once again that the entanglement entropy we are trying to compute is that of a non-local theory. It is closer in spirit to a stringy entanglement entropy, {\em not} a conventional local quantum field theory entropy. There has been some progress on computing entanglement entropies in string theory recently \cite{WittenRindler, Dabholkar}. These proceed via analytic continuations \cite{WittenRindler} of type II string orbifold partition functions \cite{Dabholkar}. But it is not yet clear if one can extract information about the natural tensor factorization structure of the Hilbert space from these seemingly formal considerations. 
One intriguing observation is that the long string limit corresponds to the $\tau_2 \rightarrow 0$ limit of  the one loop modular partition function integral, and this is where tachyonic divergences typically arise in string theories with tachyons (like the bosonic string). But the interpretation as entanglement entropy is after an analytic continuation, so it will be very interesting to develop this fully.

\section{Conclusions}

Holography is believed to be a general feature of quantum gravity, and not limited to asymptotically AdS spacetimes. The diff invariance of observables and the area law for black hole entropy, both suggest that there is a description of quantum gravity at the boundary of spacetime. In this paper, we have tried to take a few steps in understanding the nature of holography in asymptotically flat spacetimes. 

There is no sharp consensus on how to describe the dynamics of the holographic dual of flat space -- some like to describe it in terms of a celestial CFT which is co-dimension 2  \cite{RaclariuReview}, while others look for a co-dimension 1 description in terms of Carrollian theories \cite{ArjunReview}. The two descriptions must clearly be related, but the crucial question is not whether the two are equivalent, but which of these (if either) description is more {\em natural} to describe the dual. Note eg., that all the information about a conventional quantum field theory is contained in a single time slice. But for almost all questions, we find it more useful to describe things in terms of the slightly more redundant (but more intuitive) description that involves explicit time evolution. 

Given this state of affairs, it may be worthwhile taking a step back, and identifying the natural structures that arise in the hologram of flat space. In this paper, we have looked for hints about the entanglement structure of flat space, which will presumably be useful in shedding light on  identifying the correct hologram. Entanglement entropy often tells us about the useful notions of tensor factorization that are present in the system -- it has taught us about the exact locality of the boundary CFT via the RT formula, and also the approximate locality implicit in the bulk Page curve. Therefore, coming up with a {\em holographic} definition of entanglement entropy should be instructive for us in trying to guess the structural aspects of the hologram. From the bulk time foliation of flat space, it is natural to think that the entanglement entropy of flat space is best defined on its spatial boundary. In hindsight, the naturalness of our results and the richness of structure that we have found lends strong support to this possibility. 


The reason we were able to make progress on our task, is because bulk structures gave us important hints. We did not need all the details of the dual, we only needed to identify --
\begin{itemize}
\item its natural tensor factorizations, and  
\item determine the entanglement entropy of its interesting states (vacuum, dual of approximately thermal/black holes states), under these tensor factorizations. 
\end{itemize}
Interestingly, this information is precisely what one may hope to extract from an RT surface, and which is why the present project became tractable. Our task simplified to the identification of spi data that defines RT surfaces in asymptotically flat space, and became solvable.

One thing we learnt from these calculations is that bi-partite entanglement entropy of the flat space hologram has more structure than multi-partite entanglement entropy, at least in the limit where the IR cut-off has been removed. The entire spacetime is in the entanglement wedge, if we take unions of non-trivial spi-subregions. This is a distinction from AdS, where multiple non-trivial subregions  need not always be able to reconstruct the entire bulk. It is a suggestion that any two ``distinct pieces" of the flat space hologram can mutually purify each other, at least in a suitable semi-classical limit. This is suggestive of a strong form of (ultra-holographic) non-locality that is clearly important to understand better.

A second feature we will comment on before concluding is the observation that radial cut-offs are more naturally viewed as IR regulators and not as UV cut-offs in flat space. We found multiple pieces of circumstantial evidence for this. IR divergences are often viewed as more physical than UV divergences, even though this perspective is perhaps less clear for a non-local theory. 
A related point mentioned in the Introduction is that even though background subtraction has been around since \cite{Gibbons}, it has never been completely satisfactory in flat space. Also these discussions are usually in the context of entropies computed for the entire screen. Can one understand the correct way to incorporate IR divergences by looking at the explicit formulas for EE for subregions on the screen (with and without black holes) that we have provided in this paper? More generally, it seems crucial that we understand the physics  of this IR scale if we want to make conceptual progress on flat space quantum gravity. 

\section{Acknowledgments}

We thank Sam van Leuven, Jude Pereira and K. P. Yogendran for discussions. CK thanks Duy Tan University, Da Nang, and the University of the Witwatersrand, Johannesburg, for hospitality during some crucial weeks of this project. 

\section*{Appendix}
\appendix

\section{AdS, AdS Boundary and Minkowski Space}\label{AppA}

The connection between the conformal boundary of AdS and the conformal structure of flat space is useful for some of our discussions. The relation between  global, Poincare and AdS-Rindler patches and the fact that the boundary of AdS-Rindler is Rindler (the simplest ACD) is also useful to keep in mind. We review these facts to set up our coordinates and notation here. 

$\text{AdS}_{d+1}$ can be defined via the embedding ($L$ is the AdS length scale)
\begin{equation}\label{constraint}
   X_{-1}^2+X_0^2-\Vec{X}^2=L^2
\end{equation}
in $d+2$ dimensional pseudo-Euclidean space with metric
\begin{equation}
    \mathrm{d}s^2=-\mathrm{d}X_{-1}^2-\mathrm{d}X_0^2+\mathrm{d}X_1^2+...+\mathrm{d}X_d^2.
\end{equation}

Global coordinates are given by ${\displaystyle (t , r ,\theta_1 ,\cdots ,\theta _{d-2}, \phi)}$ via:
\begin{subequations}\label{eq:glob_co}
\begin{align}
    X_{-1}&=\sqrt{L^2+r^2}\cos(t/L)\\
    X_{0}&=\sqrt{L^2+r^2}\sin(t/L)\\
    \Vec{X}^2&=r^2
\end{align}
\end{subequations}
This leads to the AdS metric in global coordinates:
\begin{equation}
    ds^2=-\left(1+\frac{r^2}{L^2}\right)\mathrm{d}t^2+\frac{\mathrm{d}r^2}{\left(1+\frac{r^2}{L^2}\right)}+r^2\mathrm{d}\Omega^2_{d-1} \label{globalAdS}
\end{equation}
Here, $r\in [0,\infty), t\in (-\infty,\infty)$ and $\Omega_{d-1}$  denotes the round metric on $\mathbb{S}^{d-1}$.
The causal structure of such spaces are more conveniently represented in compactified coordinates. We define $\rho\in(0,\pi/2)$ via
\begin{equation}
    r=\tan{\rho}
\end{equation}
then the metric looks like,
\begin{equation}
    \mathrm{d}s^2=\frac{1}{\cos^2{\rho}}\left(-\mathrm{d}t^2+\mathrm{d}\rho^2+\sin^2{\rho}\mathrm{d}\Omega^2_{d-1}\right)
\end{equation}
Therefore causal structure of $\text{AdS}_{d+1}$ will be the same as that of a “solid cylinder” with metric
\begin{equation}
    \mathrm{d}s^2=-\mathrm{d}t^2+\mathrm{d}\rho^2+\sin^2{\rho}\mathrm{d}\Omega^2_{d-1}
\end{equation}
The conformal boundary has the topology of $\mathbb{R}\times\mathbb{S}^{d-1}$. 

\begin{figure}[h!]
\centering
     \begin{subfigure}[b]{0.4\textwidth}
            \centering
            \includegraphics[height=7cm]{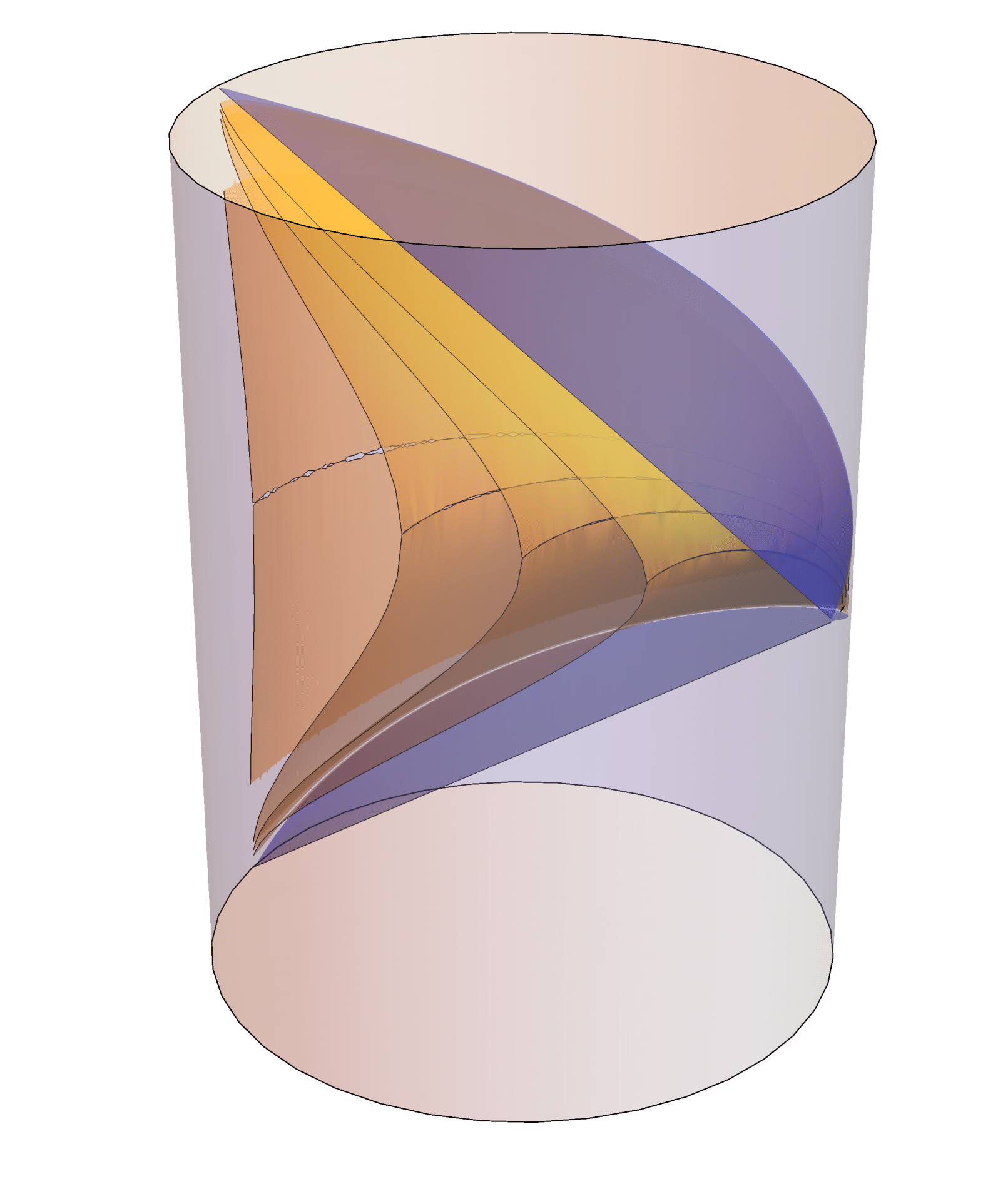}
            \caption{}
            \label{fig:poincare_a}
     \end{subfigure}
     \quad\quad\quad
      \begin{subfigure}[b]{0.4\textwidth}
            \centering
            \includegraphics[height=7cm]{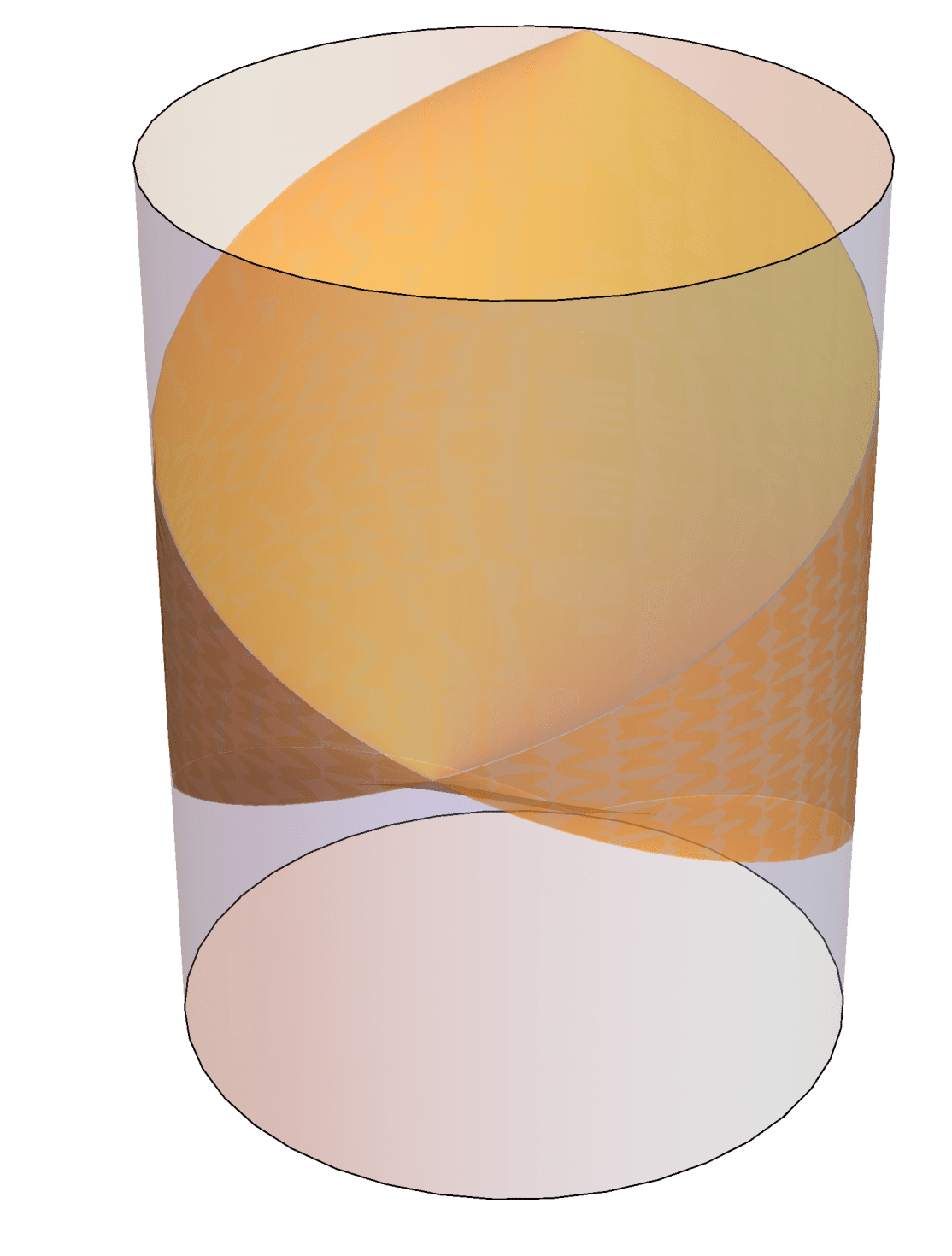}
            \caption{}
            \label{fig:poincare_b}
     \end{subfigure}  
     \caption{(a) The yellow surfaces are surfaces of constant $z$ and the blue surface is the surface for $z\rightarrow\infty$. (b) The yellow region is the full   Poincare Patch.}
\end{figure}

Poincare coordinates can be reached by the parameters $(z,x_0,x_1,...,x_{n-1})$,
\begin{subequations}
\begin{align}
    X_{-1}+X_d&=\frac{L^2}{z}\\
    X_{i}&=\frac{L}{z}x_i \quad\quad\quad\quad {i\in(0,1,...,d-1)}
\end{align}
\end{subequations}
In term of the Poincare coordinates the constraint \eqref{constraint} becomes,
\begin{equation}
    X_{-1}-X_d=z+\frac{1}{z}\left(\Vec{x}^2-x_0^2\right)
\end{equation}
Using this we can write,
\begin{subequations}\label{eq:poin_co}
\begin{align}
    X_{-1}&=\frac{1}{2}\left(\frac{L^2}{z}+z+\frac{1}{z}\left(\Vec{x}^2-x_0^2\right)\right)\label{eq:3.13}\\
    X_{i}&=\frac{L}{z}x_i \quad\quad\quad\quad {i\in(0,1,...,d-1)}\label{eq:3.14}\\
    X_d&=\frac{1}{2}\left(\frac{L^2}{z}-z-\frac{1}{z}\left(\Vec{x}^2-x_0^2\right)\right)\label{eq:3.15}
\end{align}
\end{subequations}
The coordinate ranges are $z\in(-\infty,\infty); x_i \in (-\infty,\infty)$. This yields the following metric,
\begin{equation}
    \mathrm{d}s^2=\frac{L^2}{z^2}\left(\mathrm{d}z^2+\left(\mathrm{d}\Vec{x}^2-\mathrm{d}x_0^2\right)\right)
\end{equation}
The boundary in Poincare coordinates is at $z \rightarrow 0$. The conformal boundary has the metric
\begin{equation}
     \mathrm{d}s^2=-\mathrm{d}x_0^2+\mathrm{d}\Vec{x}^2
\end{equation}
which is just Minkowski space $\mathbb{R}^{1,d-1}$. Poincare patch does not span all of AdS -- figures \ref{fig:poincare_a} and \ref{fig:poincare_b} provide plots for different $z=$const. surfaces in compactified global coordinates. 
As $z\rightarrow 0$ the surfaces approach the boundary and as $z \rightarrow \infty$ the surfaces form a wedge in the bulk. The region beyond the wedge is not accessible to Poincare coordinates. 

It should be clear that the Penrose diagram of Minkowski space is precisely the ``scarf" at the boundary of AdS Poincare patch. The compactified global coordinates $\theta$ and $t$ act as the conformal coordinates of space and time respectively on the boundary. To better understand the boundary of the Poincare patch, let us look at the geometry in $2+1$ dimensions. Then our global coordinates are spanned by $\{t, r, \theta\}$ and the Poincare patch is spanned by $\{x_0 ,x_1, z\}$. The relation between the two coordinates on the boundary is given by
\begin{subequations}\label{eq:bound_mink}
    \begin{align}
        \frac{X_1}{X_0}=\tan{\theta}=\frac{2Lx_1}{L^2-z^2+x_0^2-x_1^2}&\xrightarrow[z\rightarrow0]{}\tan{\theta}=\frac{2Lx_1}{L^2+x_0^2-x_1^2}\label{eq:3.180}\\
        \frac{X_0}{X_{-1}}=\tan{\frac{t}{L}}=\frac{2Lx_0}{L^2-z^2-x_0^2+x_1^2}&\xrightarrow[z\rightarrow0]{}\tan{\frac{t}{L}}=\frac{2Lx_0}{L^2-x_0^2+x_1^2}\label{eq:3.190}
    \end{align}
\end{subequations}
Note that these relations are the same as \eqref{eq:4.50} and \eqref{eq:4.60} with conformal factor $\Lambda=L$. This shows explicitly that the conformal boundary of the Poincare patch is the flat space Penrose diagram and the global coordinates $\theta$ and $t$ act as the space-like and time-like coordinates, respectively. It also provides another reason to think that the bulk scale in flat space is analogous to the AdS length scale and therefore is an IR scale and not a UV cut-off.
\begin{figure}[h!]
    \centering
    \includegraphics[width=7cm]{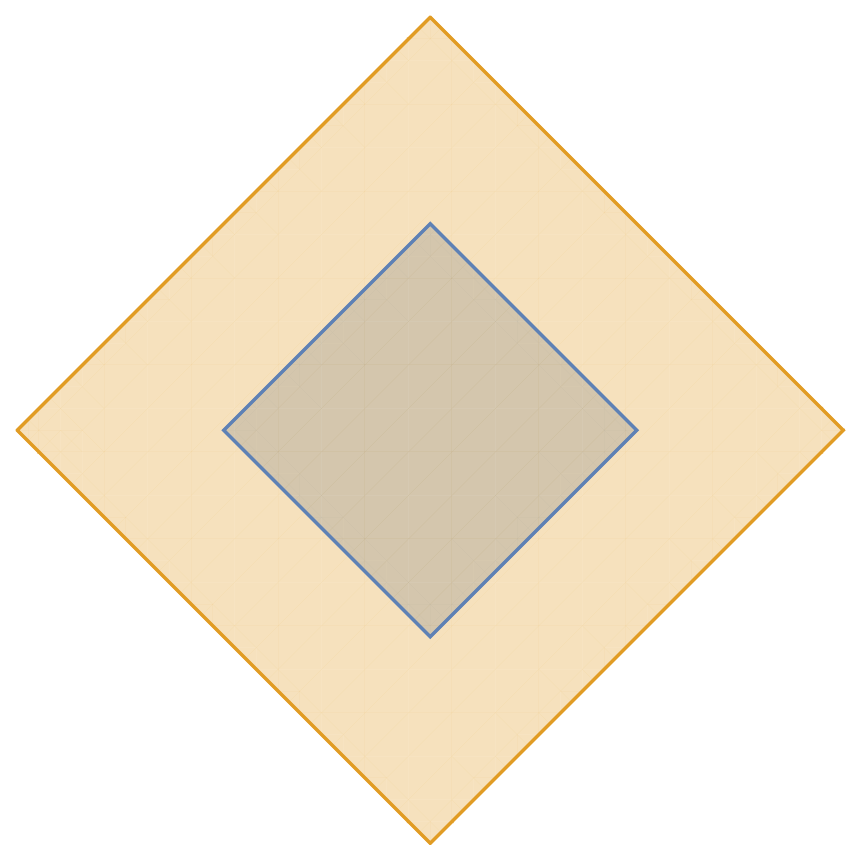}
    \caption{The boundary of the cylinder in figure \ref{fig:Rindler}. The yellow region is the Penrose diagram of flat space and the the blue region (which is the boundary of the AdS-Rindler wedge) is nothing but a causal diamond in flat space i.e. it is the domain if dependence of a spherical subregion on the boundary.}
    \label{fig:scarf}
\end{figure}

Another important set of coordinates for us are the AdS-Rindler coordinates. These are parameterized by $(\tau, \rho, u, \theta_1,...,\theta_{d-3},\phi)$,
\begin{subequations}
\begin{align}
    X_{-1}&=\rho\,\cosh{u}\\ 
    X_0&=\sqrt{\rho^2-L^2}\,\sinh{\tau/L}\\
    X_1^2+X_2^2&+...+X_{d-1}^2=\rho^2\,\sinh^2{u}\\
    X_d&=\sqrt{\rho^2-L^2}\,\cosh{\tau/L}
\end{align}
\end{subequations}
\begin{figure}[h!]
    \centering
    \includegraphics[width=6cm]{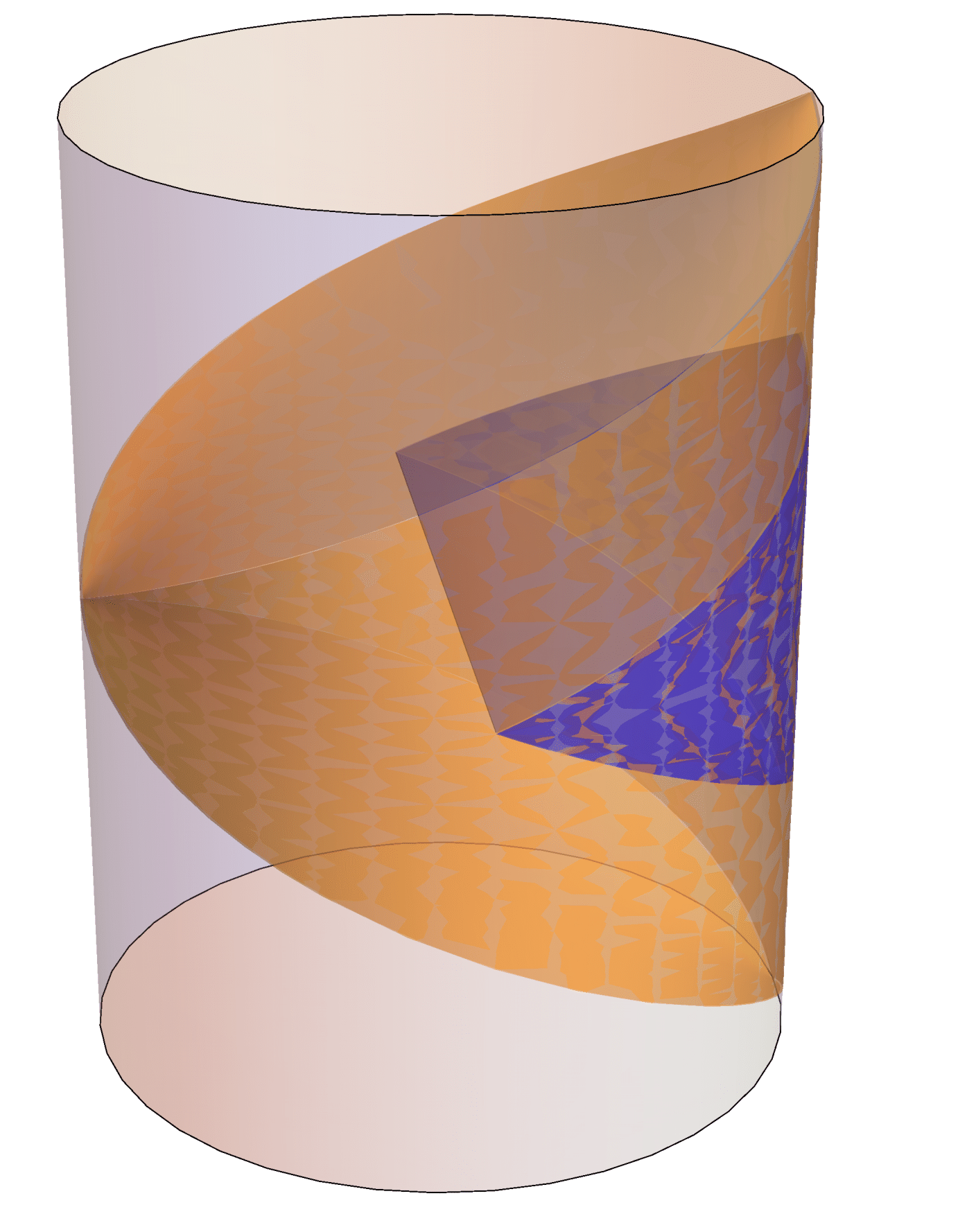}
    \caption{Poincare patch and Rindler patch presented for comparison}
    \label{fig:Rindler}
\end{figure}
The metric in these coordinates is,
\begin{equation}
    ds^2=-\left(\frac{\rho^2}{L^2}-1\right)\mathrm{d}t^2+\frac{\mathrm{d}\rho^2}{\left(\frac{\rho^2}{L^2}-1\right)}+\rho^2\left(\mathrm{d}u^2+\sinh^2{u}\,\mathrm{d}\Omega^2_{d-
    2}\right)
\end{equation}
Rindler coordinates cover an even smaller region than Poincare, and the above coordinates lie inside the Poincare patch, see Figures \ref{fig:scarf} and \ref{fig:Rindler}. It just spans the causal development of one half of the $t=0$ slice.  It has a horizon at $\rho=L$ and therefore it is sometimes called a topological black hole (it is only as much a black hole as Rindler is in flat space).

One can move the AdS-Rindler patch within the full (global) AdS spacetime by using AdS isometries. By doing this, one can in fact ensure that the boundary of AdS-Rindler overlaps precisely with the right Rindler wedge of the boundary Minkowski space of the Poincare patch, if one chooses.

\section{Solving RT Surfaces Around $\theta=0$}\label{App:RT}

In this Appendix, we will solve the RT surface equations in the form $r(\theta)$ instead of $\theta(r)$. The latter form is more adapted to the asymptotic data and was discussed in the main body of the paper, but the $r(\theta)$ form is also instructive (eg., for understanding regularity conditions from the bulk).

\subsection{Empty AdS$_4$}\label{sec:ads_4B}

The area functional associated to \eqref{globalAdS} for $d=3$ in the $r(\theta)$ form (after incorporating the symmetry of the spi-subregion) is
\begin{equation} \label{eq:area_func_2B}
    \mathcal{A}=2\pi \int \underbrace{r(\theta) \sin(\theta) \sqrt{\frac{r'(\theta)^2}{r(\theta)^2+1}+r(\theta)^2}}_\mathcal{L} \;d\theta
\end{equation}
We have set $L=1$.  The equations of motion are
\begin{equation}\label{eq:EL_eq_2B}
    \frac{\partial }{\partial r}\left(r(\theta) \sin(\theta) \sqrt{\frac{r'(\theta)^2}{r(\theta)^2+1}+r(\theta)^2}\right)=\frac{d}{d \theta}\frac{\partial }{\partial r'}\left(r(\theta) \sin(\theta) \sqrt{\frac{r'(\theta)^2}{r(\theta)^2+1}+r(\theta)^2}\right)
\end{equation}
We try a power series solution of the form
\begin{equation}\label{eq:expB}
        r(\theta)=\sum_{i=0}\;r_i\; \theta^i
\end{equation}
The expansion is around $\theta = 0$ which is the angular coordinate of the point of nearest approach to origin. Substituting \eqref{eq:expB} in \eqref{eq:EL_eq_2B} and comparing coefficients we get the solution to be:
\begin{equation}
       r(\theta)=r_0+\frac{\left(r_0^3+r_0\right)}{2} \theta ^2 +\frac{\left(9 r_0^5+14 r_0^3+5 r_0\right)}{24} \theta ^4 + ...
\end{equation}
On comparison one can see that this is an expansion for
\begin{equation}
    r = \frac{r_0 \sec{(\theta)}}{\sqrt{1+r_0^2 \tan^2{(\theta)}}}, \label{r(theta)}
\end{equation}
which also happens to be a re-writing of \eqref{eq:ads_3_sol}, with the understanding that $r_0=r_*$. 
Finally Substituting this into the Lagrangian of \eqref{eq:area_func_2B}, we get the following expansion,
\begin{align}
    \mathcal{L}=\; &  r_0^2\;\theta +\frac{ \left(8 r_0^2+5 r_0^4\right)}{6 \left(r_0^2+1\right)}\;\theta^3+\frac{\left(136 r_0^2 + 61 r_0^6 + 152 r_0^4\right)}{120 \left(r_0^2+1\right)^2}\;\theta ^5+\dots
\end{align}

\subsection{AdS$_4$ Schwarzchild}

For the AdS black hole \eqref{AdSBH} for $d = 3$, the area functional is
\begin{equation}\label{eq:area_func_4B}
    \mathcal{A}=2\pi \int \underbrace{r(\theta) \sin(\theta) \sqrt{\frac{r'(\theta)}{f_3(r)}+r^2}}_\mathcal{L} \; d\theta
\end{equation}
from which we get
\begin{equation}\label{eq:EL_eq_4B}
    \frac{\partial }{\partial r}\left(r \sin(\theta (r)) \sqrt{\frac{r'(\theta)}{f_3(r)}+r^2}\right)=\frac{d}{d \theta}\frac{\partial }{\partial r'}\left(r \sin(\theta (r)) \sqrt{\frac{r'(\theta)}{f_3(r)}+r^2}\right).
\end{equation}
With a similar method as in section \ref{sec:ads_4B}, we get the solution to be,
\begin{align}
r(\theta)=&\;\frac{1}{2} \theta ^2 \left(-r_h^3-r_h+r_0^3+r_0\right)+\frac{1}{24} \theta ^4 \left(\frac{9 r_h^6+18 r_h^4-r_h^3 r_0 \left(45 r_0^2+29\right)}{4 r_0}\right)\nonumber\\
&+\frac{1}{24} \theta ^4 \left(\frac{9 r_h^2-r_h r_0 \left(45 r_0^2+29\right)}{4 r_0}+r_0+9 r_0^5+14 r_0^3+5 r_0\right) + ...
\end{align}
On substituting this $r(\theta)$ back into the Lagrangian of \eqref{eq:area_func_4} we get,
\begin{align}
        \mathcal{L}=\; & r_0^2 \;\theta +\frac{\left(-9 r_h^3 r_0^2-9 r_h r_0^2+9 r_0^5+8 r_0^3\right)}{6 r_0}\;\theta ^3 + ...
\end{align} 

\subsection{Mink$_{4}$}

For Minkowski space with $d=3$, the area functional is
\begin{equation} \label{eq:area_func_5B}
    \mathcal{A}r=2\pi \int \underbrace{r(\theta)\sin{(\theta)}\sqrt{r^2+r'(\theta)^2}}_\mathcal{L} \; d\theta
\end{equation}
with the Euler-Lagrange equation
\begin{equation}\label{eq:EL_eq_5B}
    \frac{\partial }{\partial r}\left(r(\theta)\sin{(\theta)}\sqrt{r^2+r'(\theta)^2}\right)=\frac{d}{d \theta}\frac{\partial }{\partial r'}\left(r(\theta)\sin{(\theta)}\sqrt{r^2+r'(\theta)^2}\right).
\end{equation}
A power series solution of the form
\begin{equation}\label{eq:exp_2B}
        r(\theta)=\sum_{i=0}\;r_i\; \theta^i
\end{equation}
yields the solution
\begin{equation}\label{eq:mink_5_sol}
       r(\theta)=r_0+\frac{ r_0}{2}\;\theta ^2+\frac{5 r_0}{24}\;\theta ^4+\frac{61 r_0}{720}\;\theta ^6+\frac{277 r_0}{8064}\;\theta ^8+\frac{50521 r_0}{3628800}\;\theta ^{10}+\dots
\end{equation}
The right hand side is simply the expansion of $\frac{r_0}{\cos \theta}$, connecting with what we saw in the main text.
Substituting it in the Lagrangian, we get
\begin{equation}
    \mathcal{L}= r_0^2 \,\theta +\frac{4}{3} r_0^2 \,\theta ^3 +\frac{17}{15} r_0^2 \, \theta ^5 +\frac{248}{315} r_0^2 \,\theta ^7 + \frac{1382  r_0^2}{2835} \,\theta ^9+\dots
\end{equation}

\subsection{Mink$_5$}
The expressions take the form
\begin{equation}
    \mathcal{A}=4\pi\int \underbrace{r(\theta)^2\sin^2{(\theta)}\sqrt{r^2+r'(\theta)^2}}_\mathcal{L} \;d\theta
\end{equation}
with
\begin{equation}
        \frac{\partial }{\partial \theta} \left(r(\theta)^2\sin^2{(\theta)}\sqrt{r^2+r'(\theta)^2}\right)=\frac{d}{d r}\frac{\partial }{\partial \theta'}\left(r(\theta)^2\sin^2{(\theta)}\sqrt{r^2+r'(\theta)^2}\right)
\end{equation}
with the power series expansion
\begin{equation}
    r(\theta)=r_0+\frac{ r_0}{2}\;\theta ^2+\frac{5 r_0}{24}\;\theta ^4+\frac{61 r_0}{720}\;\theta ^6+\frac{277 r_0}{8064}\;\theta ^8+\frac{50521 r_0}{3628800}\;\theta ^{10}+\dots
\end{equation}
Which is the same as \eqref{eq:mink_5_sol} as expected of a hyperplane. Substituting it in the Lagrangian, we get,
\begin{equation}
    \mathcal{L}= r_0^3 \,\theta ^2+\frac{5}{3}  r_0^3\,\theta ^4+\frac{77}{45}r_0^3 \,\theta ^6 +\frac{88}{63}  r_0^3 \,\theta ^8+ \frac{14102 }{14175} r_0^3 \,\theta ^{10}+ ...
\end{equation}
The key observation is that the extra integration constant is not present in any dimensions in this language -- the solution is manifestly regular in the bulk.

\subsection{Schwarzchild$_4$}

For the Schwarzschild metric \eqref{Sch4}, without loss of generality we can assume the surface to be symmetric in $\phi$ direction. The area functional of a co-dimension 2 surface is
\begin{equation}
    \mathcal{A}=2\pi\int \underbrace{r \sin{\theta} \sqrt{\left(1-{\frac {r_{\mathrm {s}}}{r}}\right)^{-1}r'^{2} +r^{2}}}_\mathcal{L} \;\mathrm{d}\theta\label{eq:3.2B}
\end{equation}
and the Euler-Lagrange equation is
\begin{equation}
    \frac{\mathrm{\partial}}{\mathrm{\partial}r}\left( r \sin{\theta} \sqrt{\left(1-{\frac {r_{\mathrm {s}}}{r}}\right)^{-1}r'^{2}+r^{2}}\right)=\frac{\mathrm{d}}{\mathrm{d}\theta}\frac{\mathrm{\partial}}{\mathrm{\partial}r'}\left(r \sin{\theta} \sqrt{\left(1-{\frac {r_{\mathrm {s}}}{r}}\right)^{-1}r'^{2}+r^{2}}\right)\label{eq:el_eqB}
\end{equation}
With a power series of the form
\begin{equation}\label{eq:exp_4B}
        r(\theta)=\sum_{i=0}\;r_i\; \theta^i
\end{equation}
we get the solution
\begin{equation}
    r(\theta)=r_0+\frac{1}{2} (r_0-r_s)\;\theta ^2+\frac{1}{24}  \left(\frac{9 r_s^2-29 r_s r_0}{4 r_0}+5 r_0\right)\;\theta ^4 + \dots
\end{equation}
Substituting this into the Lagrangian of \eqref{eq:3.2B} we get,
\begin{equation}
    \mathcal{L}=r_0^2\;\theta + \frac{\left(8 r_0^2-9 r_s r_0\right)}{6}\;\theta ^3 + \dots
\end{equation}

\subsection{Schwarzchild$_5$}

For the 4+1 dimensional case the metric \eqref{Sch5} leads to the area functional 
\begin{equation}\label{eq:area_func6B}
        \mathcal{A}=4\pi \int \underbrace{r^2 \sin^2{\theta} \sqrt{\left(1-{\left(\frac {r_{\mathrm {s}}}{r}\right)^2}\right)^{-1}r'^2+r^2}}_\mathcal{L}\; d\theta
\end{equation} 
equation of motion
\begin{equation}
    \frac{\mathrm{\partial}}{\mathrm{\partial}r}\left( r^2 \sin^2{\theta_1} \sqrt{\left(1-{\frac {r^2_{\mathrm {s}}}{r^2}}\right)^{-1}r'^{2}+r^{2}}\right)=\frac{\mathrm{d}}{\mathrm{d}\theta}\frac{\mathrm{\partial}}{\mathrm{\partial}r'}\left(r^2 \sin^2{\theta_1} \sqrt{\left(1-{\frac {r^2_{\mathrm {s}}}{r^2}}\right)^{-1}r'^{2}+r^{2}}\right) 
\end{equation}
with a power series solution 
\begin{equation}
    r(\theta)=\; r_0+\frac{1}{2} \theta ^2 \left(r_0-\frac{h^2}{r_0}\right)+\frac{1}{24} \theta ^4 \left(5 r_0-\frac{3 h^4+22 h^2 r_0^2}{5 r_0^3}\right)+...
\end{equation}
On substituting this $r(\theta)$ back into the Lagrangian of \eqref{eq:area_func6B} we get

\begin{align}
    \mathcal{L}= r_0^3 \theta ^2 +\frac{ \left(10 r_0^4 -9 h^2 r_0^2-5 h r_0^3\right)}{6 (h+r_0)} \theta ^4 + ...
\end{align}

\section{RT Surfaces in Cylindrical Coordinates}\label{cylinder}

We have worked with spherical polar coordinates in most of this paper, partly because comparisons with black hole and AdS results is easiest in that language. But when solving for flat space RT surfaces in these coordinates, we found that the structure of integration constants was a bit strange. With a power series expansion in powers of $1/r$, we found two integration constants in all dimensions other than 3+1 (in 2+1 the extra integration constant was trivial, but it was there). The fact that the extremal surface equations are second order ODEs suggests two integration constants, one of which is to be eliminated via a bulk condition. So the absence of the second integration constant in 3+1 dimensions has to be explained. We will do this by solving the same system in cylindrical polar coordinates, where the system is exactly solvable in general dimensions, and finding that the extra constant in 3+1 dimensions corresponds to a log term. In 4+1 dimensions, the solution will also provide us a sanity check of our $1/r$ expansion where it reproduces both the constants after transforming to  polar coordinates. Cylindrical coordinates are also useful in showing that the solution can be found in closed form. 

The metric of the $t=0$ slice of flat space in $d$-dimensional cylindrical coordinates can be expressed as follows:
\begin{equation}
    ds^2=dz^2+d\rho^2+\rho^2d\Omega_{d-2}^2 
\end{equation}
Assuming a surface symmetric about the pole ($z$-axis), which can be written as $z(\rho)$, the area functional is given by:
\begin{equation}
    \mathcal{A} =\frac{2\pi^{\frac{\pi-1}{2}}}{\Gamma\left(\frac{\pi-1}{2}\right)} \int \underbrace{\rho^{d-2}\sqrt{z'(\rho)+\rho^2}}_{\mathcal{L}}\; d\rho
\end{equation}
By solving the Euler-Lagrange equations, we find that the general solution in cylindrical coordinates is of the form:
\begin{equation}
    z = i \rho \;\; _2F_1\left(\frac{1}{2},\frac{1}{2(d-2)};1+\frac{1}{2(d-2)};\frac{\rho^{2(d-2)}}{b^2}\right) + c
\end{equation}
Here, $_2F_1$ represents a hypergeometric function.

\subsection{3+1D}

\begin{figure}[h!]
    \centering
    \includegraphics[width=10cm]{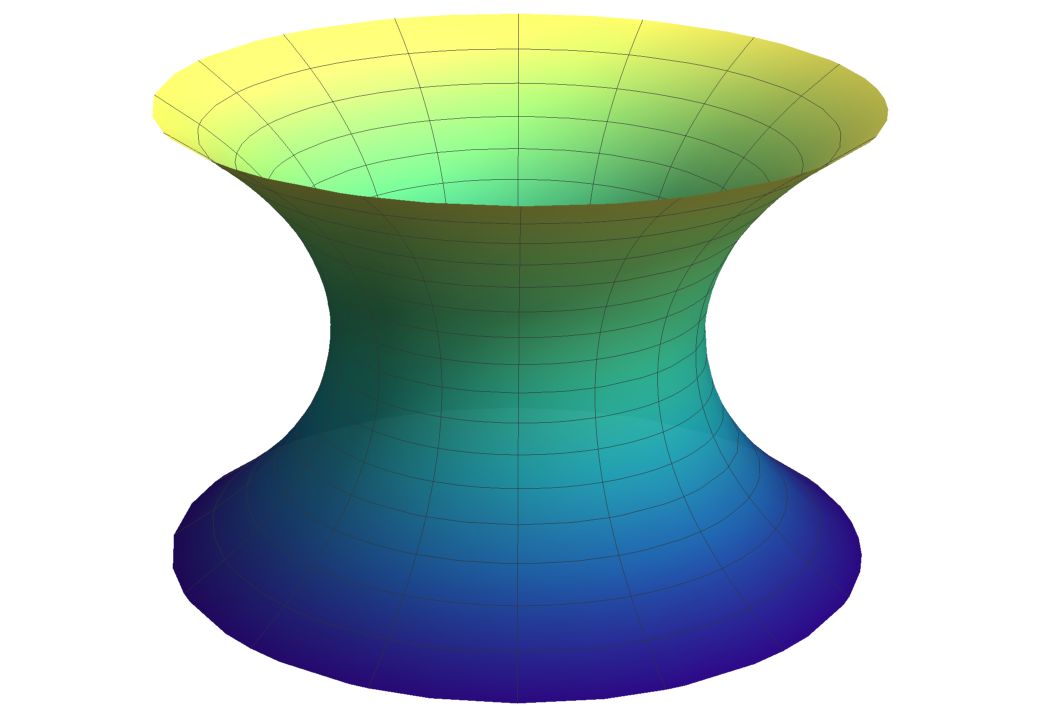}
    \caption{Catenoid}
    \label{fig:catenoid}
\end{figure}

For a geometry where $d=3$, this becomes
\begin{equation}\label{eq:catenoid}
    z=b\ln{(\sqrt{\rho^2-b^2}+\rho)} +c
\end{equation}
and can be rearranged to get
\begin{equation}
    \rho = b \cosh{\left[\frac{z}{b}-\left(\frac{c}{b}+\ln{b}\right)\right]}
\end{equation}
with $c=-b\ln{b}$, we obtain the equation of a catenoid (refer to Figure \ref{fig:catenoid})
\begin{equation}
    \rho =b\cosh {\left(\frac {z}{b}\right)}
\end{equation}
For $b = 0$, we get a plane:
\begin{equation}
    z=c
\end{equation}
Expanding \eqref{eq:catenoid} for large $\rho$, we have:
\begin{equation}
    z = c + b \log \left(\frac{1}{\rho }\right) +\frac{b^3}{4}\frac{1}{\rho ^2} +\frac{3 b^5}{32}\frac{1}{\rho ^4} + \frac{5 b^7}{96}\frac{1}{ \rho ^6} + \dots
\end{equation}
From this, we observe that the Catenoid is what is responsible for the logarithmic fall off. Therefore, the $\theta(r)$ expansion attempted in Section \ref{mink4-curve} does not represent the most general solution since it did not incorporate a logarithmic fall off. But it {\em is} the solution that we seek, because our bulk condition eliminates the catenoid.

\subsection{4+1D}

In $d=4$, the most general surface is given by:
\begin{equation} \label{eq:4D_minimal}
    z = i \rho \;\; _2F_1\left(\frac{1}{2},\frac{1}{4};\frac{5}{4};\frac{\rho^{4}}{b^2}\right) + c
\end{equation}
Expanding \eqref{eq:4D_minimal} for large $\rho$, we get:
\begin{equation}
    z = c  - \frac{4 b \Gamma \left(\frac{5}{4}\right)}{ \Gamma \left(\frac{1}{4}\right)} \frac{1}{\rho} - \frac{2 b^3 \Gamma \left(\frac{5}{4}\right)}{5 \Gamma \left(\frac{1}{4}\right)}\frac{1}{\rho^5} + \dots
\end{equation}
Interestingly, if we write \eqref{eq:4D_minimal} in spherical polar coordinates ($z=r \cos{(\theta)}$, $\rho = r \sin{(\theta)}$), substitute \eqref{eq:exp_2}, and expand for large radius, we recover \eqref{eq:mink_5sol}. Therefore, we observe that the extra constant in dimensions higher than 3+1 is visible already in an expansion in $1/r$. Essentially, the catenoid-like solution in higher dimensions does not lead to a log.

\bibliographystyle{JHEP}
\bibliography{bib}
\end{document}